\documentclass[12pt]{elsarticle}
\makeatletter
\def\ps@pprintTitle{%
 \let\@oddhead\@empty
 \let\@evenhead\@empty
 \def\@oddfoot{\centerline{\thepage}}%
 \def\@evenfoot{\thepage}
 \let\@evenfoot\@oddfoot}
\makeatother
\usepackage[final]{changes} 

\usepackage{microtype}
\usepackage[margin=1in]{geometry}
\usepackage{amsmath,amssymb,amsthm,bm}
\usepackage{subcaption}
\biboptions{sort&compress}
\usepackage{epstopdf}
\usepackage{float}
\usepackage{xcolor}
\usepackage[labelfont=bf]{caption}
\usepackage[export]{adjustbox}
\AppendGraphicsExtensions{.tif}

\definecolor{C0}{HTML}{1F77B4}
\definecolor{C1}{HTML}{FF7F0E}
\definechangesauthor[color=C0]{R1} 
\definechangesauthor[color=C1]{R2} 
\definechangesauthor[color=C3]{RA} 

\definecolor{C1}{HTML}{FF7F0E}
\definecolor{C2}{HTML}{2ca02c}
\definecolor{C3}{HTML}{d62728}
\definecolor{C4}{HTML}{9467bd}
\definecolor{C5}{HTML}{8c564b}

\begin{document}

\title{Block structured adaptive mesh refinement and strong form elasticity approach to phase field fracture with applications to delamination, crack branching and crack deflection}
\author[auburn]{Vinamra Agrawal\corref{cor1}}
\author[uccs]{Brandon Runnels}
\cortext[cor1]{Corresponding author: vinagr@auburn.edu}
\address[auburn]{Department of Aerospace Engineering, Auburn University, Auburn, AL, USA}
\address[uccs]{Department of Mechanical and Aerospace Engineering, University of Colorado, Colorado Springs, CO USA}
\begin{abstract}
    Fracture is a ubiquitous phenomenon in  most composite engineering structures, and is often the responsible mechanism for catastrophic failure.
    Over the past several decades, many approaches have emerged to model and predict crack failure.
    The phase field method for fracture uses a surrogate damage field to model crack propagation, eliminating the arduous need for explicit crack meshing.
    In this work a novel numerical framework is proposed for implementing hybrid phase field fracture in heterogeneous materials.
    The proposed method is based on the ``reflux-free'' method for solving, in strong form, the equations of linear elasticity on a block-structured adaptive mesh refinement (BSAMR) mesh.
    The use of BSAMR enables highly efficient and scalable regridding, facilitates the use of temporal subcycling for explicit time integration, and allows for ultra-high refinement at crack boundaries with minimal computational cost.
    The method is applied to a variety of simple heterogeneous structures: laminates, wavy interfaces, and circular inclusions.
    In each case a non-dimensionalized parameter study is performed to identify regions of behavior, varying both the geometry of the problem and the relative fracture energy release rate.
    In the laminate and wavy interface cases, regions of delamination and fracture correspond to simple analytical predictions.
    For the circular inclusions, the modulus ratio of the inclusion is varied as well as the delamination energy release rate and the problem geometry.
    In this case, a wide variety of behaviors was observed, including deflection, splitting, delamination, and pure fracture.
\end{abstract}

\begin{keyword}
    Fracture mechanics, phase field fracture, delamination, composites, block-structured adaptive mesh refinement, strong form elasticity
\end{keyword}

\maketitle
\section{Introduction}

    Crack propagation is known to be the prominent cause of failure of composite engineering structures ranging from catastrophic structural failure to aging and/or catastrophic failure of solid rocket propellants \cite{hu2017experimentally,kadiresh2008experimental,wang2020tensile}.
    Composite engineering materials - including structural (such as nanolayered Cu-Nb or carbon-reinforced plastic) and energetic (such as AP+HTPB or Aluminum+PVDF solid propellant) - can exhibit an even wider range of fracture behavior due to the combination of delamination, fracture, and crack deflection resulting from material heterogeneity. 
    Enhanced numerical methods are required in order to predict fracture behavior in such materials, and to create materials that are failure-resistant by design.
    
    While crack propagation in homogeneous materials have been extensively studied, heterogeneous materials have received relatively less attention.
    In heterogeneous materials, crack can undergo deflection, repeated arrest and nucleation depending on the material architecture and constituent properties.
    This is especially important when considering (micro-) crack propagation in the micro and mesoscales.
    Micro-cracks can interact with material interfaces, pores and grain boundaries leading to complex crack paths \cite{kumar2007crack}.
    Recently crack propagation in heterogeneous materials have been studied from the context of effective fracture toughness \cite{hossain2014effective,hsueh2018stress}.
    Other works have focused on predicting crack paths and failure in composites \cite{espadas2019phase,zhang2019phase,tan2020phase} and bio-inspired materials \cite{murali2011role,khaderi2014failure,singh2019interplay}.
    Insights into crack paths and failure mechanisms can lead to designing failure resistant materials.
    The interaction of crack with material interfaces can generally lead to either crack arrest, penetration through the interface, or redirection along the interface leading to delamination. 
    This behavior is the result of energetically competing mechanisms involving mechanical work and the cost of creating new surfaces.

    Understanding and modeling of crack propagation in materials has been a topic of significant study since the early 70s; since then computational fracture mechanics has become a field of study in its own right.
    Computational fracture mechanics methods generally fall into one of two categories, which we term discrete boundary methods or diffuse boundary methods.
    Of the discrete boundary methods there are two main approaches.
    The first is the extended finite element method (XFEM), in which the classical finite element method is enriched with specialty elements specifically designed for the purpose of capturing singularities at crack tips.
    Another discrete boundary method is the scaled boundary finite element method (SBFEM).
    SBFEM is a dimensional reduction technique, in which the problem domain is reduced to the boundary of the solid, and for which the solution can be ``scaled'' to the crack tip analytically.
    This enables the highly efficient solution of many types of crack problems, but is relatively limited in its scope of crack types \cite{zhang2019scaled,hell2019enriched,jiang2019modelling}.
    A full review of these and other discrete element methods is outside the scope of this present work; we refer the reader to \cite{egger2019discrete,sedmak2018computational} for a comprehensive review.
    Although both types of discrete boundary methods have enjoyed considerable success, they are both limited by the topologically cumbersome task of explicit tracking of crack fronts for complex crack patterns.

    Diffuse boundary methods, or ``phase field methods'' aim to address some of the discrete boundary tracking issues by tracking the crack path using a smoothly varying scalar damage field \cite{ambati2015review}.
    The phase field damage variable is rooted in a rigorous, variational approach to fracture, from which differential equations for the phase field evolution are derived  \cite{francfort1998revisiting}.
    Phase field methods \replaced[id=R2,comment={2.1}]{implicitly}{effortlessly} account for highly complex fracture patterns, and can naturally predict crack initiation as well as branching/merging \cite{bourdin2000numerical,bourdin2007numerical,bourdin2008variational}. 
    The use of a diffuse interface requires the introduction of a length scale parameter $\xi$ for the crack, which is then used to regularize the underlying energy functional \cite{francfort1998revisiting}.
    It has been shown that the variational method is consistent with linear elastic fracture mechanics (LEFM), and it has been extended to study brittle fracture \cite{bourdin2007numerical, kuhn2010continuum, miehe2015phasea}, dynamic fracture \cite{borden2012phase,schluter2014phase,karma2001phase,hofacker2012continuum}, anisotropic materials \cite{li2015phase,teichtmeister2017phase}, heterogeneous materials \cite{nguyen2016phase,hansen2020phase}, functionally graded materials \added[id=R1,comment={1.1}]{\cite{doan2016hybrid,dinachandra2020phase,kumar2021phase,natarajan2019phase}} interface strength \cite{schneider2016phase}, crack nucleation \cite{tanne2018crack,brach2019phase,kumar2020revisiting}, ductile fracture \cite{ambati2015phase,miehe2015phaseb,kuhn2016phase} and fatigue induced crack propagation \cite{mesgarnejad2019phase}. 
    Efforts have also been directed towards understanding the role of damage degradation function and the damage energy penalty term on the solutions \cite{lorentz2011gradient,pham2011gradient,pham2011issues,pham2013onset,kuhn2015degradation,wu2018length}.
    Recently, the phase field formulation of fracture mechanics has been modified to incorporate cohesive zone elements to account for interfacial strength \cite{carollo2018modeling, tarafder2020finite, quintanas2020phase}.
    While these methods have successfully modeled interface failure \cite{roy2017phase, carollo2018modeling, dhas2018phase, sehr2019interface, quintanas2020phase}, there is a need for a systematic understanding of the competing mechanisms driving crack behavior.
    This is especially important for designing conventional fiber-reinforced composites and novel materials such as bio-inspired sutured interface materials \cite{li2013generalized, cao2019experimental, liu2020interfacial}.
    
    The main limitation of diffuse interface methods is their computational cost.
    The mesh must be fine enough to capture the gradient of the damage field, and therefore the requisite number of points must scale with the inverse of $\xi$.
    Given that $\xi\to0$ to recover the exact solution, this computational cost can become prohibitive.
    One way to avoid this probitive cost is to use spectral methods, employing the fast Fourier transforms (FFT)  for improved memory efficiency \cite{chen2019fft,ernesti2020fast}.
    An alternative way to allay this challenge is to take advantage of the fact that the grid need not be high-resolution everywhere, but only near the crack itself, and that in the far field, the grid can often be extremely coarse.
    To achieve this, one might ``pre-refine'' the grid in anticipation of where the crack is expected to go -- but to do so would be to undo the crack-path-agnostic advantage of the phase field method.
    A better alternative is to alter the mesh in response to the crack propagation, using adaptive mesh refinement (AMR).
    Progress on AMR for phase field fracture has progressed recently on a number of fronts.
    The discontinuous Galerkin method (DG) with the finite element method has been used to implement single-level AMR with phase field \cite{muixi2020hybridizable}. 
    Alternately, the finite cell method has been used together with h and p refinement to achieve multiple levels of refinement on a regular grid \cite{nagaraja2019phase}.
    Other recent attempts have been made to hybridize the phase field method with XFEM by restricting the solution area to the crack tip \cite{giovanardi2017hybrid,patil2018local}.
    
    \begin{figure}
      \centering
      \includegraphics[width=14cm]{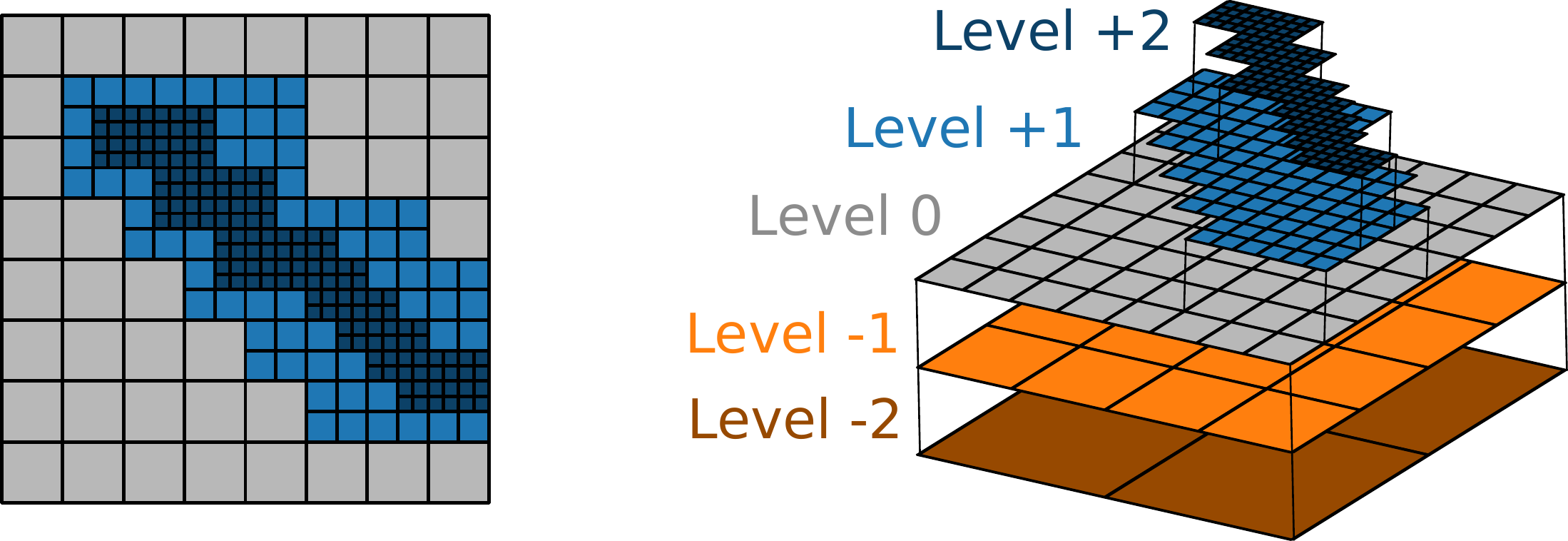}
      \caption{Schematic of BSAMR multi-level refinement scheme}
      \label{fig:multilevel}
    \end{figure}

    Block-structured AMR (BSAMR) is an alternative type of AMR that has enjoyed substantial popularity in the high performance computational fluid mechanics community \cite{zhange2019amrex,hittinger2013block,schornbaum2018extreme,dubey2014survey}, but has been relatively absent in solid mechanics applications.
    In a BSAMR mesh, each refinement level is stored independently (Figure~\ref{fig:multilevel}).
    No connectivity information is required, which avoids incurring additional communication overhead and improves parallel performance.
    For explicit time integration, each level is evolved independently, with in-level communication happening via ghost cells.
    This independent time integration enables temporal subcycling, where fine levels are allowed to evolve with a smaller timestap than coarse levels, avoiding overly restrictive CFL conditions resulting from the fine grid.
    Fine-grid information is then averaged down to coarse grids, and updated coarse grid information is communicated to the fine grid using ghost cells.
    For implicit solves, the AMR multi-level grid structure can be used naturally for geometric multigrid by adding additional coarse grids below, and implementing the usual interpolation and restriction operators.
    In recent work, a strong-form elasticity solver was developed in conjunction with the reflux-free method for coarse-fine syncronization to enable the solution of elasticity problems on a BSAMR grid.
    The reflux-free method eliminates the need for explicit handling of the coarse-fine transition region, without impacting the convergence of the solver \cite{runnels2020massively}.
    In this work we apply the BSAMR strong form elasticity solver, along with the implementation of hybrid phase field fracture mechanics on a BSAMR grid, to achieve a highly scalable phase field fracture framework.

    This paper is organized as follows.
    In section \ref{sec:PFFOverview}, we provide a brief overview of the hybrid phase field fracture including equilibrium equations.
    We describe the FD based reflux-free BSAMR solver in section \ref{sec:Alamo}, followed by three examples in section \ref{sec:Examples}: a crack impinging on a planar surface, delamination/fracture of a wavy interface, and crack interaction with a circular inclusion.
    
\section{Hybrid model for phase field fracture} \label{sec:PFFOverview}
    We use the hybrid formulation of the phase field fracture proposed by Ambati {\it et al}.
\cite{ambati2015review} as a general reference for this section.
    The smooth crack field is represented by $c(x)$ , such that $c=0$ inside the crack and $1$ outside.
    The hybrid formulation combines the tension-compression symmetric (TC-symmetric, sometimes called {\it ``isotropic''}) \cite{bourdin2000numerical} and tension-compression asymmetric (TC-asymmetric, sometimes called {\it ``anisotropic''}) formulations \cite{miehe2015phasea} for improved computational efficiency.
    The TC-symmetric energy functional \cite{bourdin2000numerical} is given by
    \begin{equation}
        \mathcal{F}_i(\bm{u},c) = \int_\Omega \left(c^2+\eta\right) \Psi_0(\bm{\varepsilon}) dx + \int_\Omega G_c \Big[\frac{\left(1-c\right)^2}{4\xi} + \xi|\nabla c|^2\Big] dx,
    \end{equation}
    where $\eta \ll 1$ for computational stability, $G_c$ is the material fracture toughness, $\bm{u}$ is the displacement field, $\bm{\varepsilon} = \text{sym}\,\text{grad}\,\bm{u}$ is the strain field and $\Psi_0$ is the elastic strain energy density.
    For a linear elastic isotropic material with Lam\'{e} constants $\lambda$ and $\mu$, the elastic strain energy density is given by
    \begin{equation}
        \Psi_0(\bm{\varepsilon}) = \frac{1}{2}\lambda \left(\text{tr}\, \bm{\varepsilon}\right)^2 + \mu \text{tr}\left(\bm{\varepsilon}^2\right).
    \end{equation}
    The Euler Lagrange equilibrium equations result in
    \begin{align}
        \text{div}\,\bm{\sigma} &= \bm{0}, \quad
        \bm{\sigma} = \left(c^2+\eta\right) \frac{\partial \Psi_0}{\partial \bm{\varepsilon}}, \nonumber\\
        0 &= 2c\mathcal{H} - G_c \left[2\xi \Delta c + \frac{1-s}{2\xi}\right],
    \end{align}
    where $\mathcal{H} := \max_{\tau \in [0,t]} \Psi_0(\bm{\varepsilon}(x,t))$.
    
    In order to capture TC-asymmetry, it is necessary to decompose the stress tensor in such a way as to prevent crack growth under compression.
    Literature has predominantly focused on either the deviatoric-volumetric split \cite{schluter2014phase}, or the spectral decomposition \cite{miehe2015phasea,borden2012phase}.
    In this work we adopt the latter.

    \added[id=R2,comment={2.3}]{
        It should be noted that there are some well-known issues with the proposed spectral splitting scheme under multi-axial loading \cite{sun2021poro}.
        It does not explicitly take the crack orientation into account which can lead to the violation of crack boundary conditions \cite{strobl2015novel}.
        Other splitting schemes that aim to address these shortcomings include that recently proposed by Lo {\it et al} \cite{lo2019phase}, as well as the ``directional split'' \cite{steinke2019phase}.
        Application of these more advanced techniques is not the present focus, but will constitute future work.
    }

    We assume an additive decomposition of the elastic strain energy $\Psi_0 = \Psi_0^+ + \Psi_0^-$, which uses the spectral decomposition of the strain tensor $\bm{\varepsilon} = \sum_{i=1}^{d} \varepsilon_i \hat{\bm{n}}_i\otimes \hat{\bm{n}}_i$, where $\varepsilon_i$ are principal strains, $\hat{\bm{n}}_i$ are principal directions and $d$ is the dimension. 
    Finally,
    \begin{equation}
        \Psi_0^\pm = \frac{1}{2}\lambda \left(\text{tr}\, \bm{\varepsilon_\pm}\right)^2 + \mu \text{tr}\left(\bm{\varepsilon}_\pm^2\right), \quad 
        \bm{\varepsilon}_\pm = \sum_{i=1}^{d} (\varepsilon_i)_\pm \hat{\bm{n}}_i\otimes \hat{\bm{n}}_i
    \end{equation}
    The regularized TC-asymmetric energy functional is given by
    \begin{equation}
        \mathcal{F}_a(\bm{u},c) = \int_\Omega \left[\left(c^2+\eta\right) \Psi_0^+ + \Psi_0^-\right] dx + \int_\Omega G_c \left[\frac{\left(1-c\right)^2}{4\xi} + \xi|\nabla c|^2\right] dx.
    \end{equation}
    The Euler Lagrange equilibrium equation for the TC-asymmetric formulation are given by
    \begin{align}
        \text{div}\,\bm{\sigma} &= \bm{0}, \quad
        \bm{\sigma} = \left(c^2+\eta\right) \frac{\partial \Psi_0^+}{\partial \bm{\varepsilon}} + \frac{\partial \Psi_0^-}{\partial \bm{\varepsilon}}, \nonumber\\
        0 &= 2c\mathcal{H}^+ - G_c \left[2\xi \Delta c + \frac{1-s}{2\xi}\right],
    \end{align}
    where $\mathcal{H}^+ := \max_{\tau \in [0,t]} \Psi_0^+(\bm{\varepsilon}(x,t))$.
    
    In this work, the hybrid model is used to leverage the computational efficiency of the TC-symmetric model, while still retaining a crack driven by tensile component of the elastic energy. 
    The equilibrium equations of the hybrid model are given by
    \begin{subequations}
    \begin{align}
        \text{div}\,\bm{\sigma} &= \bm{0}, \quad
        \bm{\sigma} = \left(c^2+\eta\right) \frac{\partial \Psi_0}{\partial \bm{\varepsilon}}, \label{eq:HybridModelEquations_equilibrium}\\
        0 &= 2c\mathcal{H}^+ - G_c \left[2\xi \Delta c + \frac{1-s}{2\xi}\right],  \label{eq:HybridModelEquations_crackeq}\\
        \forall \bm{x}&: \Psi_0^+ < \Psi_0^- \Rightarrow c = 1,
        \label{eq:HybridModelEquations_ineq}
    \end{align}
    \end{subequations}
    where (\ref{eq:HybridModelEquations_ineq}) is added to prevent crack interpenetration. 
    In this work, we replace the crack equilibrium equation (\ref{eq:HybridModelEquations_crackeq}) with a Ginzburg-Landau type evolution law \cite{kuhn2008phase}:
    \begin{equation}
        \dot{c} = -M\left[2c\mathcal{H}^+ - G_c \left(2\xi \Delta c + \frac{1-s}{2\xi}\right)\right], \label{eq:HybridLandauEquation}
    \end{equation}
    where $M\geq 0$ is the mobility parameter associated with crack propagation.

    In general, phase field fracture problems are solved using a staggered approach where displacement and crack fields are solved alternately \cite{miehe2010phase}, or in a monolithic fashion \cite{farrell2017linear,jodlbauer2020matrix} where both fields are solved simultaneously.
    Monolithic solvers are generally more efficient; for instance, recent work has demonstrated the efficacy of the BFGS algorithm in the context of phase field \cite{wu2020bfgs}.
    However, staggered approaches have been shown to be  sufficiently robust \cite{miehe2010phase}, and because of their accuracy and relative ease of implementation, they are commonly used for phase field fracture.
    In this framework we also adopt a staggered method, both to maintain consistency with the large body of existing literature, and also to take advantage of the temporal subcycling offered by BSAMR.
    Details on the elasticity solver are discussed in the following section.
    Between each elastic solve, the crack field is allowed to evolve until it stagnates, so that $\dot{c}=0$.
    Therefore, while (\ref{eq:HybridLandauEquation}) could be interpreted as a kinetic law, we use it here merely as a gradient descent method, for which the ``mobility'' acts as a step size. 
    Consequently, there is no dependence on $M$ beyond requiring that it be small enough to not violate the CFL condition.
    Thus, we leave the exploration of transient crack phenomena to future work.

\section{Reflux-free Finite Difference Solver} \label{sec:Alamo}

\begin{figure}
  \centering
  \includegraphics[width=0.32\linewidth,clip,trim=7.2cm 5.5cm 1.7cm 3.6cm]{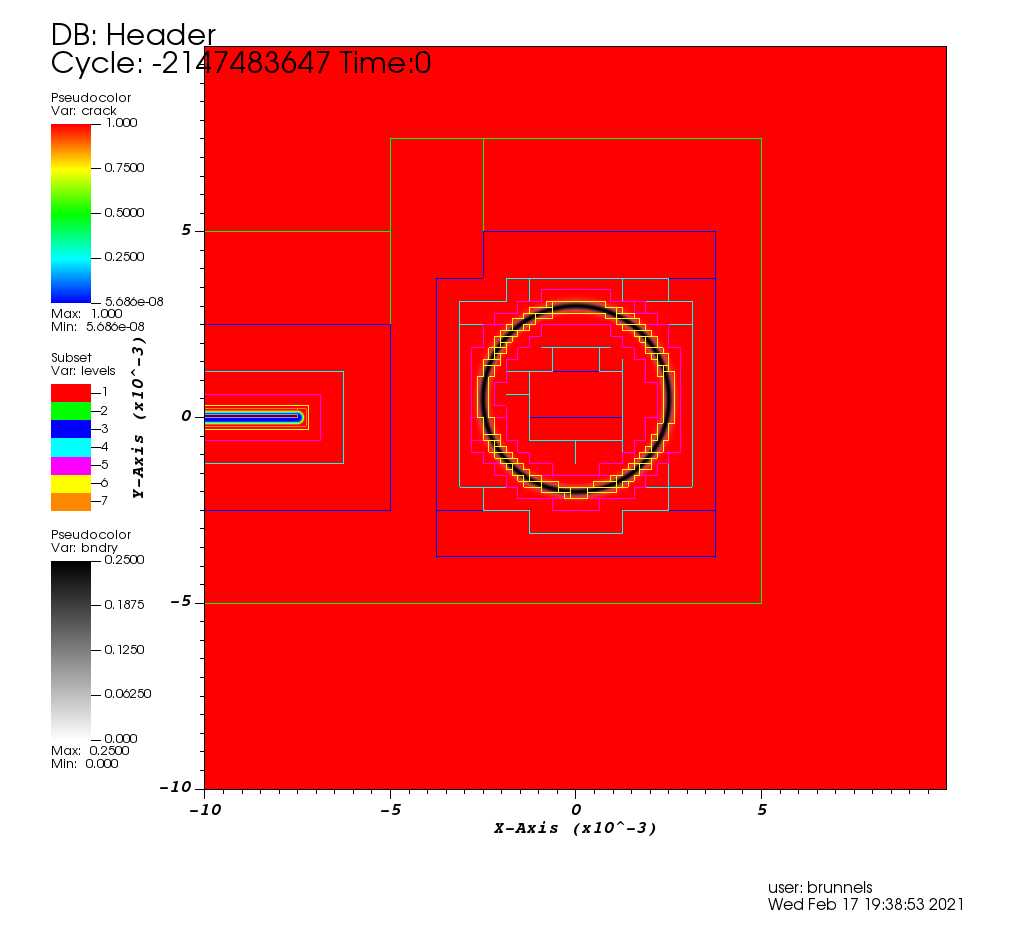} %
  \includegraphics[width=0.32\linewidth,clip,trim=7.2cm 5.5cm 1.7cm 3.6cm]{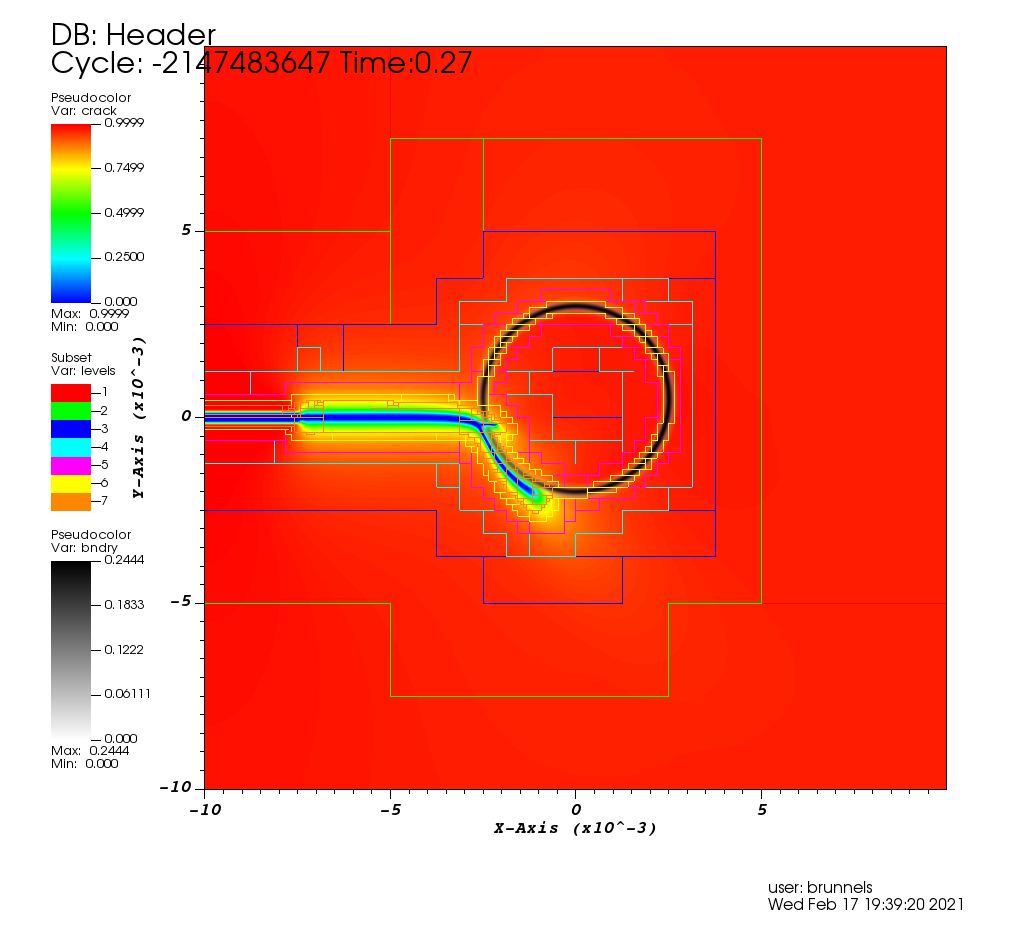} %
  \includegraphics[width=0.32\linewidth,clip,trim=7.2cm 5.5cm 1.7cm 3.6cm]{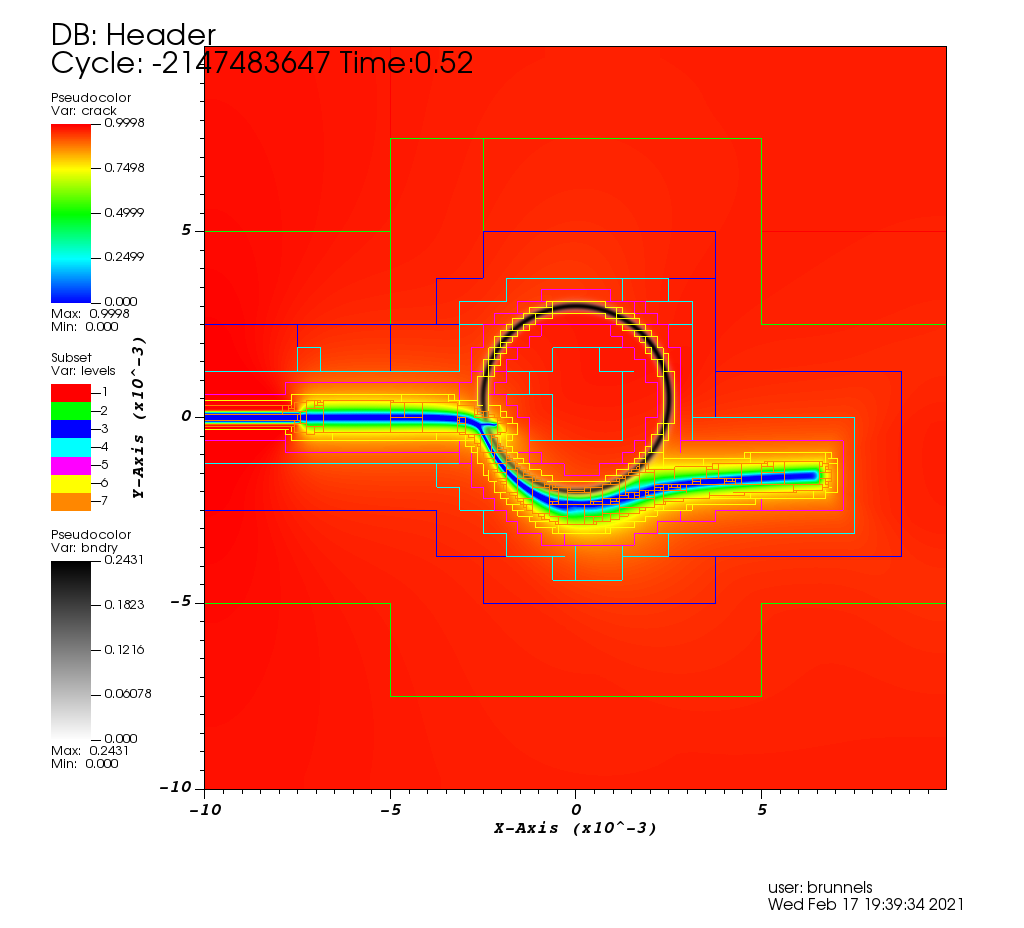} 
  \caption{
    Block structured adaptive mesh refinement (with seven total levels) of crack propagation into and around a circular inclusion.
    \replaced[id=R2]{The mesh is initially refined around material heterogeneities, and then dynamically evolves throughout the simulation to capture the moving crack tip.}{Mesh refinement is used to track the crack front as well as to resolve the material interface.}
  }\label{fig:bsamr}
\end{figure}


Fracture mechanics problems, by definition, contain highly localized features (cracks) in an otherwise relatively uninteresting ambient medium.
As a result, strategic meshing is required to provide adequate resolution at the features of interest without incurring excessive computational cost due to the large problem size.
The majority of previous phase field fracture work uses the finite element method to solve the equations of elasticity on a mesh that is either refined {\it a priori} \cite{borden2012phase} or using adaptive mesh refinement \cite{ambati2015review,heister2015primal,borden2016phase}. 

The phase field method for fracture, unlike many other methods such as the cohesive zone model, does not require any explicit meshing scheme.
Indeed, phase field methods often perform optimally when implemented on a regular grid, rather than a specifically tailored unstructured mesh.
However, they are often implemented using such a mesh anyway, in order to accommodate the requisite FEM elasticity solver.
As a result, such numerical schemes carry the computational burden of FEM based adaptive mesh refinement, while offering few benefits to the implementation of the phase field fracture model itself.

In this work, we employ an alternative numerical implementation that is optimal for both the implicit and explicit calculations in phase field fracture.
Block-structured AMR (BSAMR) is a specific variety of adaptive mesh refinement that partitions the mesh into individual levels, where each level is associated with a specific degree of refinement (Figure~\ref{fig:bsamr}).
Each level is broken up into individual cubic regions called patches, where each patch contains data stored as a regular, structured grid.
Each patch, in turn, is stored using a space-filling Z-curve scheme in order to maximize data locality.
Over the course of temporal integration, patches are evolved independently from one another, then eventually synced up using averaging between levels.
The multi-level structure enables efficient temporal subcycling, allowing finer levels to evolve with a different timestep than the coarser levels.
Though not explored specifically in this work, temporal subcycling is highly advantageous for fourth-order fracture evolution models, allowing for the timestep to decrease commensurate with the increase of spatial discretization, per the CFL condition.
In the BSAMR method, mesh refinement and coarsening occur at regular intervals (see section below) and trigger the regeneration of grids using the Berger-Rigoutsos algorithm \cite{berger1991algorithm}.
This scheme offers myriad advantages, including an ability to scale easily to very large platforms, the ready portability to heterogeneous architectures such as GPUs, and the localization of data for the most numerically intensive operations.

Despite its many advantages, however, BSAMR has seen very limited application to solid mechanics and elasticity problems.
This is primarily due to the inherent difficulties in solving the equations of static elasticity on a BSAMR mesh.
In this work, we build on previous contributions by the authors toward the development of a strong form elasticity solver \cite{runnels2020massively} for numerical methods using BSAMR.
In this method, the equations of elasticity along with boundary conditions are posed as a single partitioned operator $D:\Omega\times\mathbb{R}^n\to\mathbb{R}^n$ on problem domain $\Omega$, defined as
\begin{align}
  D(\bm{x})(\bm{u}) =
  \begin{cases}
    \operatorname{div}\big[\mathbb{C}(\bm{x})\big(\operatorname{grad}\bm{u} - \bm{\varepsilon}^0(\bm{x})\big)\big] & \bm{x}\in\operatorname{int}(\Omega) \\
    \bm{u} & \bm{x}\in\partial_1\Omega \ \ \text{(Dirichlet boundary)}\\
    \mathbb{C}(\bm{x})\big(\operatorname{grad}\bm{u} - \bm{\varepsilon}^0(\bm{x})\big)\, \bm{n}(\bm{x}) & \bm{x}\in\partial_2\Omega \ \ \text{(Traction boundary)}\\
    (\operatorname{grad}\bm{u})\, \bm{n}(\bm{x}) & \bm{x}\in\partial_3\Omega \ \ \text{(Neumann boundary)}
  \end{cases}\label{eq:strongform}
\end{align}
where $\mathbb{C}$ is the spatially varying modulus tensor, $\mathbb{\varepsilon}^0$ is a spatially varying eigenstrain tensor, and $\bm{n}$ is the outward-facing normal vector.
It should be noted that there is an important distinction to be made in the strong form between the $\partial_2\Omega$ (Traction) boundary and the $\partial_3\Omega$ (Neumman) boundary. In the homogeneous and eigenstrain-free case, commonly used to impose symmetry conditions on linear materials, they are clearly equivalent. However, in cases where the Neumann condition is inhomogeneous (less common) or when the eigenstrain is nonzero (more common), they require distinct implementations.
The standard equations of elasticity, along with boundary conditions, are thus expressible as
\begin{align}\label{eq:goveq}
  D(\bm{x})(\bm{u}) = \bm{r}(\bm{x}),
\end{align}
where the right-hand side term $\bm{r}$ is defined piecewise as 
\begin{align}
  \bm{r}(\bm{x}) =
  \begin{cases}
    -\bm{b}(\bm{x}) & \bm{x}\in\operatorname{int}(\Omega) \\
    \bm{u}^0(\bm{x}) & \bm{x}\in\partial_1\Omega \\
    \bm{t}^0(\bm{x}) & \bm{x}\in\partial_2\Omega \\
    \bm{q}^0(\bm{x}) & \bm{x}\in\partial_3\Omega,
  \end{cases}
\end{align}
with $\bm{b}$ the body force, $\bm{u}^0$ the prescribed displacement, $\bm{t}^0$ the prescribed traction, and $\bm{q}^0$ the prescribed gradient of $\bm{u}$ normal to the boundary.
The latter is used in this work for imposing symmetry conditions by setting $\bm{q}^0$ to zero normal to the line of symmetry, and combining with a Dirichlet condition in the direction normal to displacement. Traction displacements could be used instead to achieve this same effect, but Neumann boundaries offer a slight performance advantage in the strong form.

Operator (\ref{eq:strongform}) is implemented numerically in strong form, where $\mathbb{C}(\bm{x})$ describes the spatial variation of $\mathbb{C}$ in accordance with (\ref{eq:HybridModelEquations_equilibrium}).
The traditional difficulties of a strong form implementation are avoided by the use of C++ templating and operator overloading, allowing the highly efficient evaluation of fourth-order matrix multiplication operations.
Crucially, the use of strong form allows for the definition of level-consistent numerical operators $D_{n}$, such that the constants associated with $D_n$ may be interpolated and restricted between levels in a manner consistent with the numerical discretization.
That is, if $I$ is the injective interpolation operator and $R$ the surjective restriction operator, the implementation of $D_{n},D_{n+1}$ must be such that the following are maintained:
\begin{align}\label{eq:consistency}
  D_{n} = R\circ D_{n+1}\circ I 
\end{align}
for all $n$.
(Note that the inverse relation is generally not true.
A discussion of this is included in \ref{sec:multi_level_consistency}.)
It is worthwhile to mention, at this point, a pitfall that can arise when solving near-singular elasticity problems, such as phase field fracture, with highly localized boundaries:
It is sometimes the case that operators $\ldots,D_{-1},D_{0},D_{1},\ldots$ are initialized separately, such that (\ref{eq:consistency}) does not hold.
Such an inconsistency can cause substantial convergence issues, even though the separate initialization may actually be more accurate.
To avoid such problems, coarse operators should always be initialized by $D_{n-1}=R\,D$ in all the areas of overlap, to ensure consistency.

BSAMR is highly conducive to the use of multigrid for solving (\ref{eq:goveq}), as the refinement levels posess the same structure as coarsening levels in the geometric multigrid method.
$D_0,D_1,\ldots$ correspond to the discretized operator on the unrefined level, the first level of refinement, and so on.
$D_{-1}$ corresponds to the first coarsened multigrid level, $D_{-2}$ to the second coarsening level, and so on.
The operators $R,I$ for AMR levels are the same as those used for the multigrid interpolation and restriction operations.

The method of successive over-relaxation is used for the multigrid smoothing operation, defined here to be
\begin{align}
  S = (D^{D} + \omega D^{L})^{-1}\Big(\omega \bm{r} [\omega D^{U} + (\omega-1)D^{D}]\bm{x}\Big),
\end{align}
where $D^{D}$, $D^{L}$, $D^{U}$ are the diagonal, lower, and upper components of $D$.
It was determined that a relaxation factor of $\omega=0.08$ produced optimal convergence for the problems discussed here.
For the bottom-level linear solve, the smoothing operator $S$ was used instead of the more commonly used stabilized biconjugate gradient method, due to the near-singular behavior of the operator.

The implementation of the restriction operator across the coarse/fine boundary is one of the key difficulties in solving elasticity using BSAMR.
This work builds on the previous development of the ``reflux-free'' algorithm, which eliminates the need for an explicit coarse/fine treatment by using additional ghost nodes at the boundary.
For a full discussion of the reflux-free method, the reader is referred to \cite{runnels2020massively}.

\section{Examples} \label{sec:Examples}

We use an in-house code (Alamo) to solve crack equations associated with the hybrid formulation of phase field fracture to systematically study the interaction of crack with material interfaces in two dimensions.
We study three cases under Mode-I loading: interaction of crack with angled planar interface, interaction of crack with wavy interface, and crack propagation through a maze of spherical inclusions.
To focus on the interplay between driving forces governing delamination and fracture, we consider simplified problems that are representative examples of crack/interface interaction.
Each material is represented by a phase variable $\phi_i$, while the crack field is represented by $c$.
For all cases, we choose a square domain $x,y\in [-0.01,0.01]\times[-0.01,0.01]$ with $64\times 64$ base mesh, with six levels of BSAMR refinement.

Mesh regridding occurred every 10 Level-0 timesteps.
Refinement occurred until the following criterion was met:
\begin{equation}\label{eq:refinement_criterion}
    \max(|\nabla \phi|, |\nabla c|)|\Delta \bm{x}^n| > 0.1 
    \ \ \ \forall n \in (0,N), \ \ \ \forall \bm{x}^n\in\Omega^n.
\end{equation}
In the above, $n$ indicates mesh level with a maximum of $N$ possible levels, $\bm{x}^n$ the position vector within each level's domain $\Omega^n$, and $\Delta \bm{x}^n$ the mesh resolution vector for level $n$.
\added[id=R2,comment={2.2}]{The $\nabla c$ term induces the regridding dependence on the material interfaces, which are static in time and trigger the initial mesh adaption. The $\nabla c$ term, which changes in time, causes the mesh to dynamically refine with the crack tip, to guarantee adequate crack tip refinement regardless of the path that the crack chooses to take.
Grid points for which \ref{eq:refinement_criterion} does not hold are coarsened, although this does not generally happen because the gradient of $c$ never decreases. 
}

All materials (matrix and inclusion) are linear elastic isotropic with same Lam\'{e} parameters $\lambda = 121.15\times 10^9$, $\mu = 80.77\times 10^9$ \added[id=RA]{unless indicated otherwise}.
Materials have same fracture energy release rate $G_c = G_{cf} = 2.7\times 10^3$, crack length scale $\xi = 10^{-5}$ and mobility $M=10^{-5}$. 
We choose $\eta = 5\times 10^{-4}$ for numerical stability.
In all the examples outlined below, the following boundary conditions are used.
\begin{align}
    \bm{u} = 0, &\quad y=-0.01,\;\forall\, x\in[-0.01,0.01] \nonumber \\
    \bm{u} = \bm{u}_0, &\quad y=0.01,\;\forall\, x\in[-0.01,0.01] \nonumber\\
    \bm{u}\cdot \hat{\bm{e}}_1 = 0, &\quad x = \pm 0.01,\;\forall\, y\in[-0.01,0.01] \nonumber \\
    \left(\bm{\sigma}\hat{\bm{e}}_1\right)\cdot \hat{\bm{e}}_2 = 0, &\quad x = \pm 0.01,\;\forall\, y\in[-0.01,0.01].
\end{align}
Finally, the material interface in the examples presented below is diffused with a length scale $b = 5\times 10^{-4}$ to ensure a large enough $b/\xi$ ratio to avoid re-scaling the interfacial delamination energy release rate \cite{hansen2019phase}.
To compute the overall fracture energy release rate, we computed the total crack volume in the domain and the total elastic energy $\left(c^2 + \eta\right)\Psi_0(\bm{\varepsilon})$ at every step.
We then smoothed the data using a Savitzky-Golay filter, before differentiating it to compute the incremental energy release rate.

\subsection{Angled planar interface}

\begin{figure}[h]
  \begin{minipage}[t][6cm][t]{\linewidth}
    \begin{minipage}[t][6cm][t]{0.3\linewidth}
      \includegraphics[width=\linewidth,clip,trim=0cm -1.3cm 0cm 0cm]{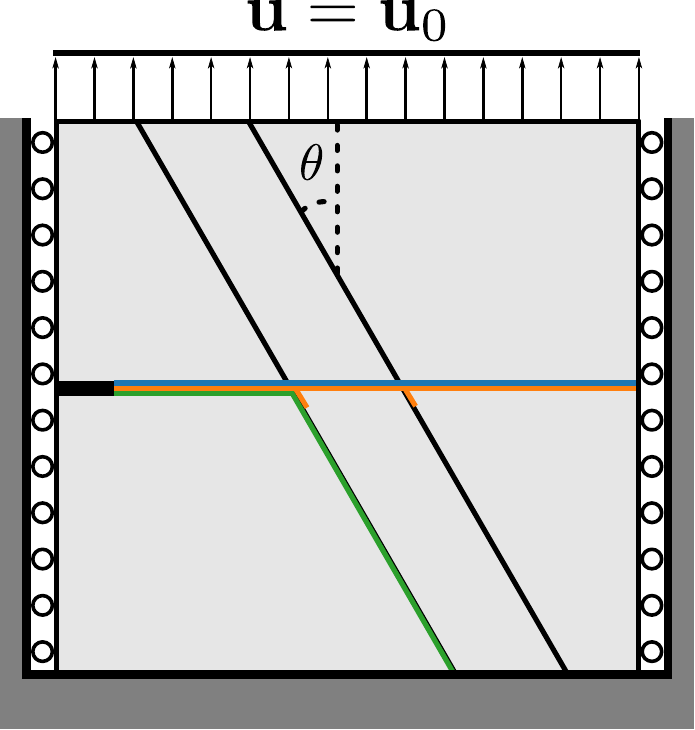}
    \end{minipage}\hfill
    \begin{minipage}[t][6cm][t]{0.6\linewidth}
      \includegraphics[width=\linewidth]{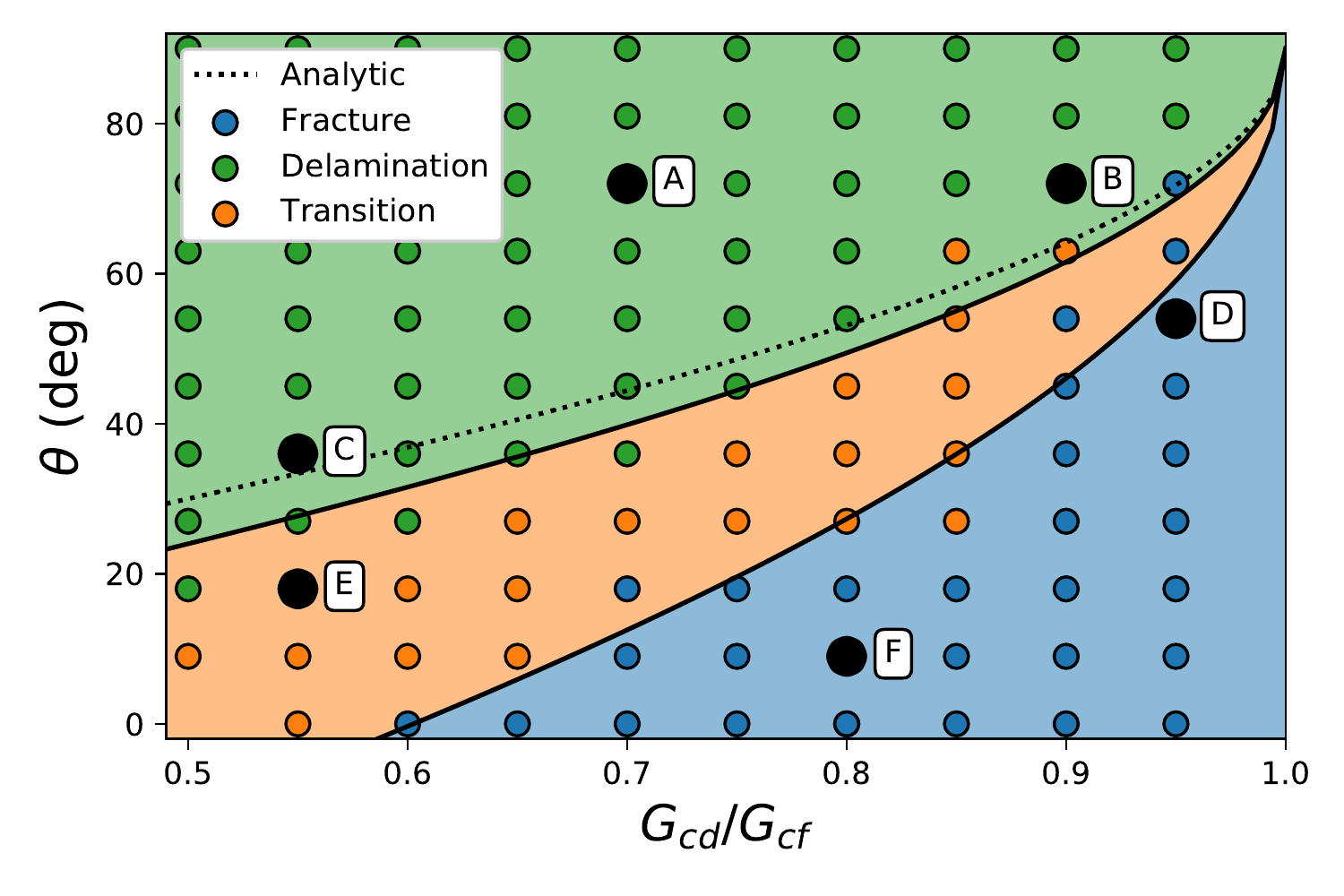}
    \end{minipage}
  \end{minipage}
  \caption{
    Laminate: 
    ({\bf Left})
    Mode I loading of a material with twin interfaces angled at $\theta$ from the vertical axis.
    Colored lines indicate sample crack patterns: the {\color{C0}\bf blue} line indicates pure fracture, {\color{C2}\bf green} indicates maximum delamination, and {\color{C1}\bf yellow} indicates the transition region.
    ({\bf Right})
    Paramater survey of delamination and fracture for an angled planar interface.
    Callouts A-F correspond to visualizations of crack/delamination patterns in Figure~\ref{fig:planar_laminate_visualizations}, and colors correspond to the convention in the Right figure.
    Contours are a visual aid to assist in identifying regions of behavior.
  }
  \label{fig:angled_interface}
\end{figure}

\begin{figure}[h]
  \centering
  \begin{minipage}{0.15\linewidth}
    \includegraphics{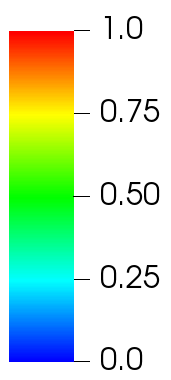}
  \end{minipage}
  \begin{minipage}{0.75\linewidth}
  \begin{minipage}{\linewidth}
    \begin{subfigure}{0.3\linewidth}\includegraphics[width=\linewidth,clip,trim=7.2cm 5.5cm 1.7cm 3.5cm,cfbox=C2 3pt 0pt]{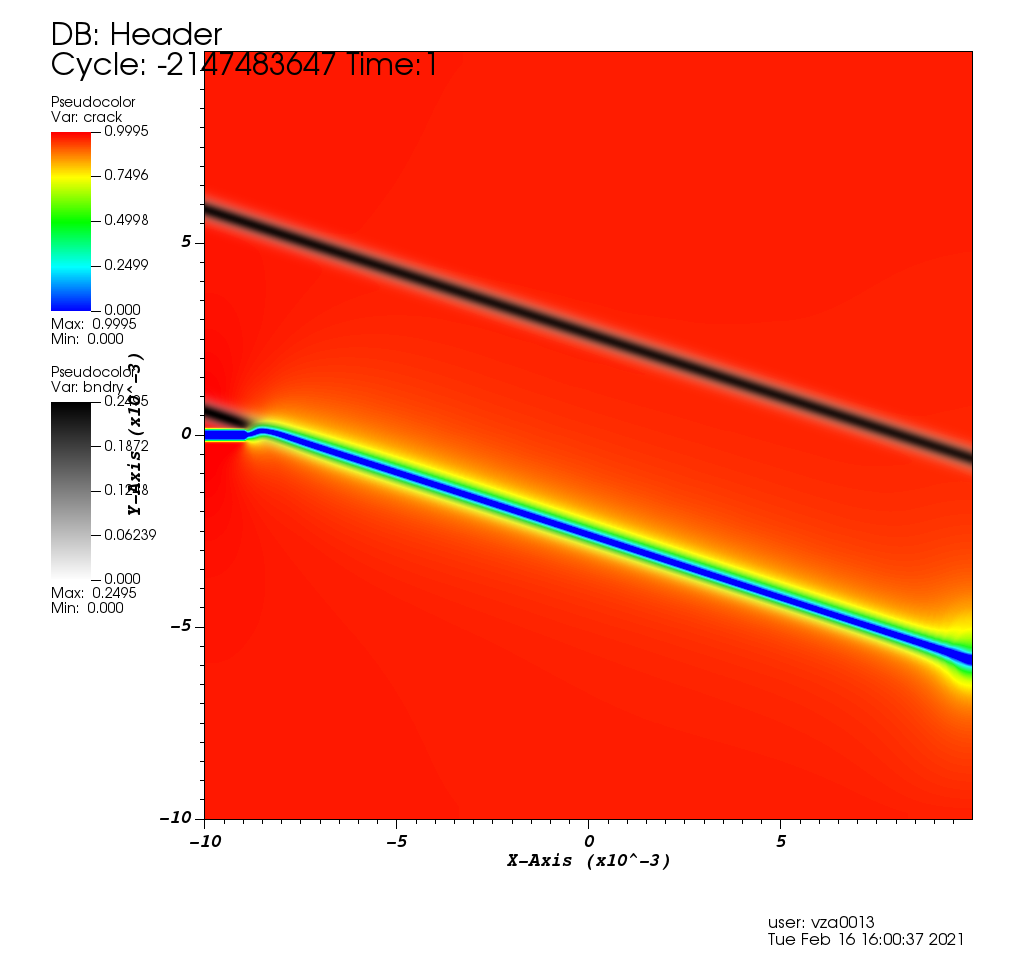} \caption{}\label{fig:laminate_output_1035}\end{subfigure}\hspace{10pt}%
    \begin{subfigure}{0.3\linewidth}\includegraphics[width=\linewidth,clip,trim=7.2cm 5.5cm 1.7cm 3.5cm,cfbox=C1 3pt 0pt]{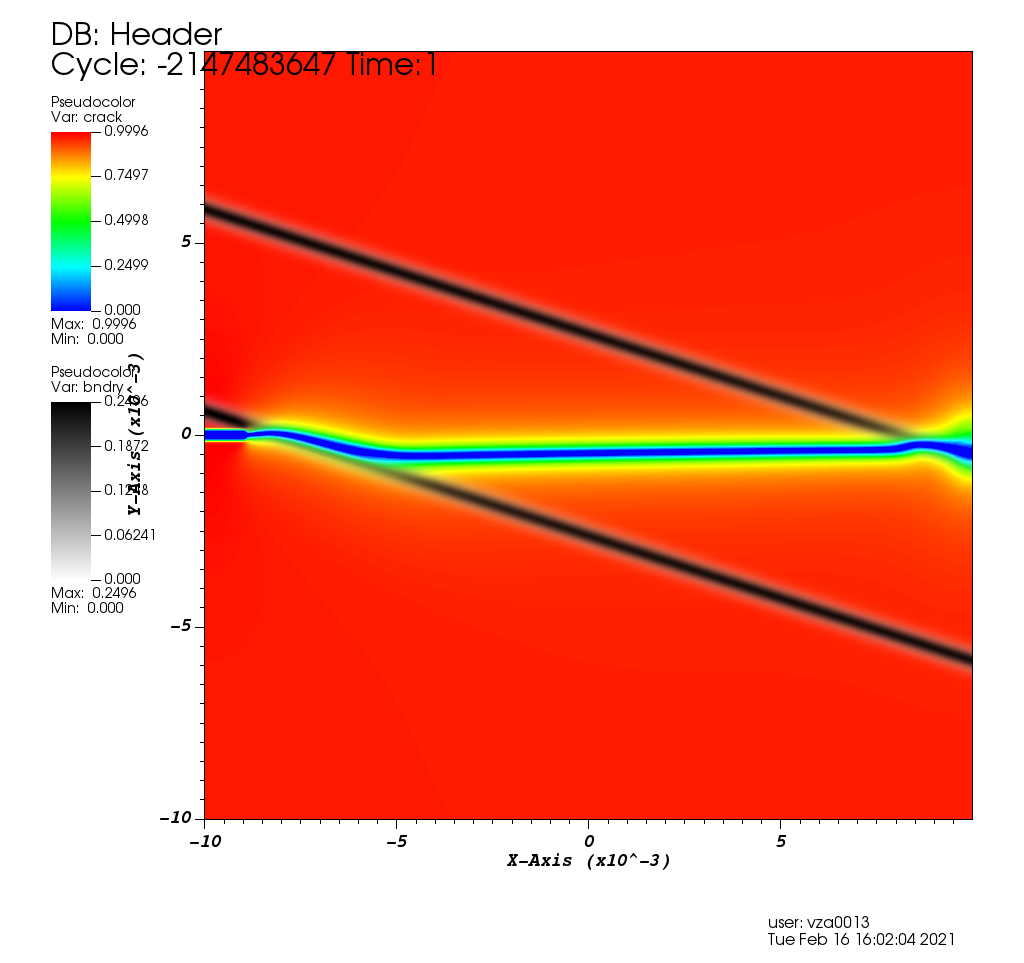} \caption{}\label{fig:laminate_output_1037}\end{subfigure}\hspace{10pt}%
    \begin{subfigure}{0.3\linewidth}\includegraphics[width=\linewidth,clip,trim=7.2cm 5.5cm 1.7cm 3.5cm,cfbox=C2 3pt 0pt]{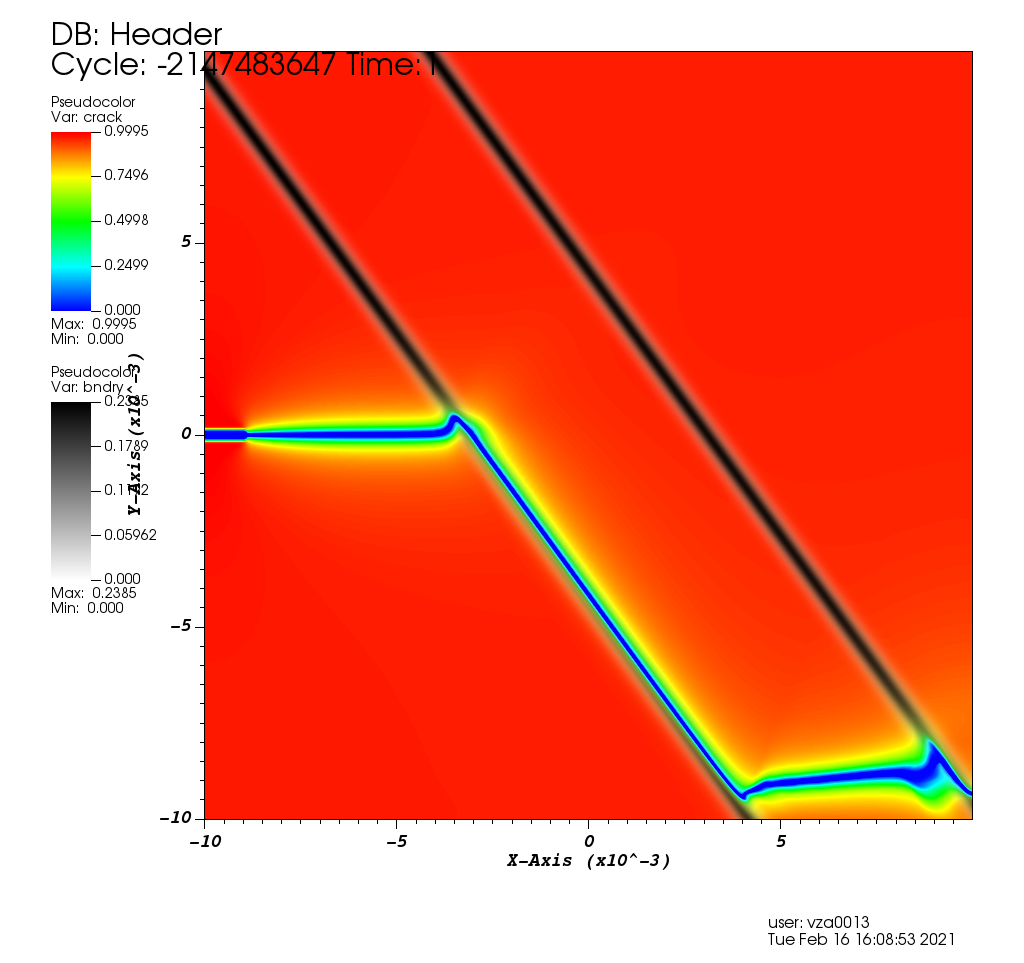} \caption{}\label{fig:laminate_output_1776}\end{subfigure}
  \end{minipage}
  \begin{minipage}{\linewidth}
    \begin{subfigure}{0.3\linewidth}\includegraphics[width=\linewidth,clip,trim=7.2cm 5.5cm 1.7cm 3.5cm,cfbox=C0 3pt 0pt]{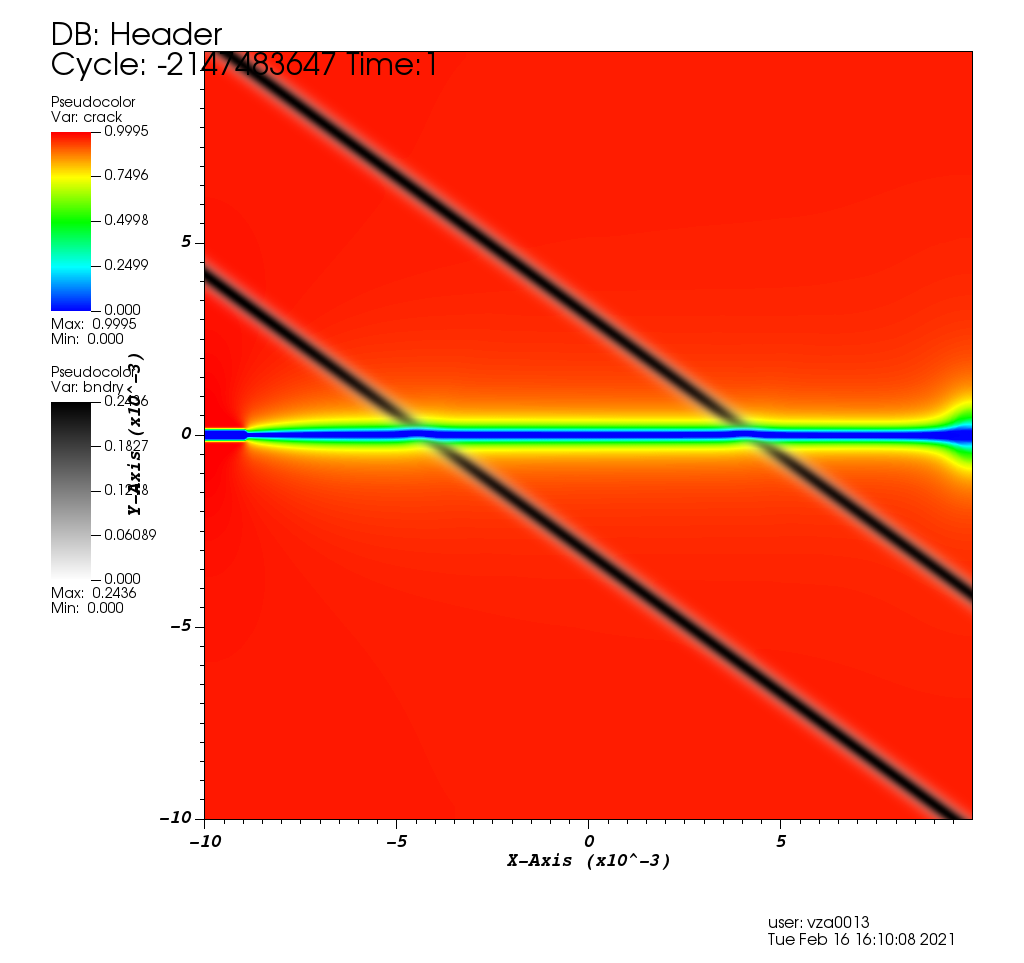} \caption{}\label{fig:laminate_output_1790}\end{subfigure}\hspace{10pt}%
    \begin{subfigure}{0.3\linewidth}\includegraphics[width=\linewidth,clip,trim=7.2cm 5.5cm 1.7cm 3.5cm,cfbox=C1 3pt 0pt]{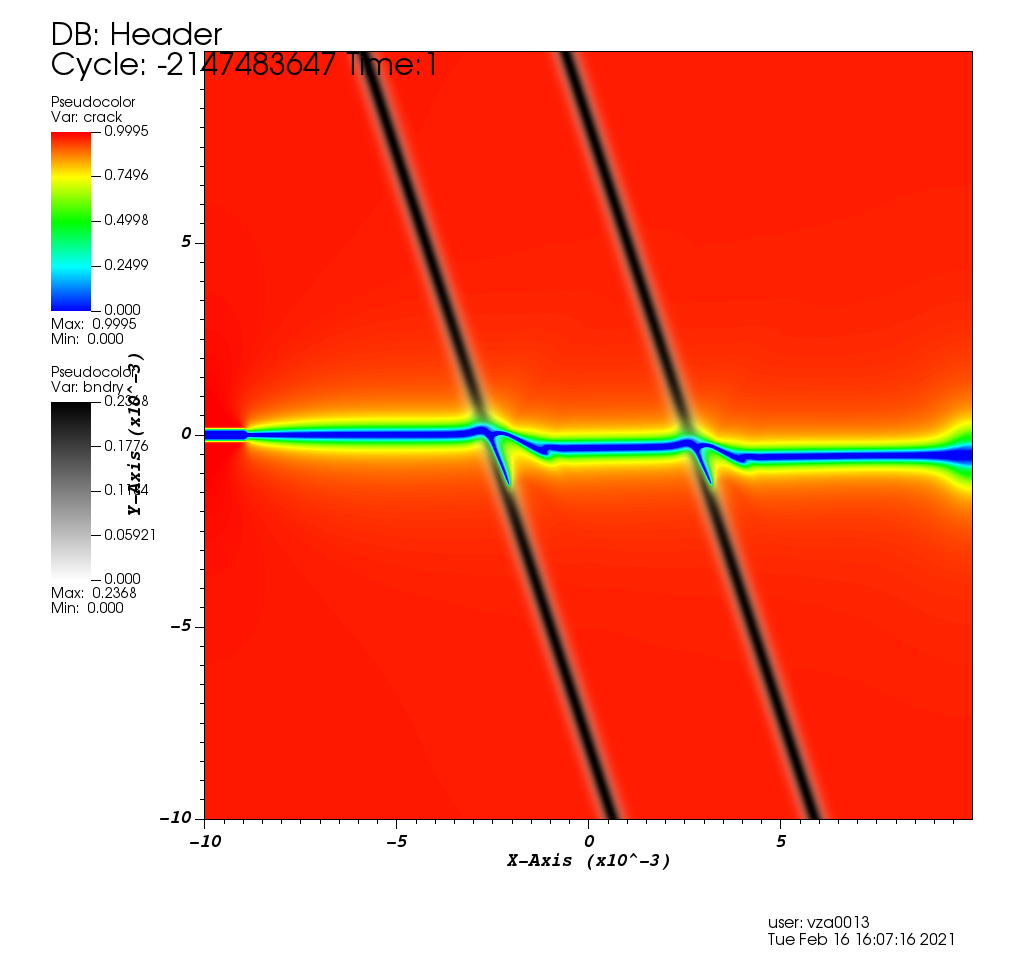} \caption{}\label{fig:laminate_output_1766}\end{subfigure}\hspace{10pt}%
    \begin{subfigure}{0.3\linewidth}\includegraphics[width=\linewidth,clip,trim=7.2cm 5.5cm 1.7cm 3.5cm,cfbox=C0 3pt 0pt]{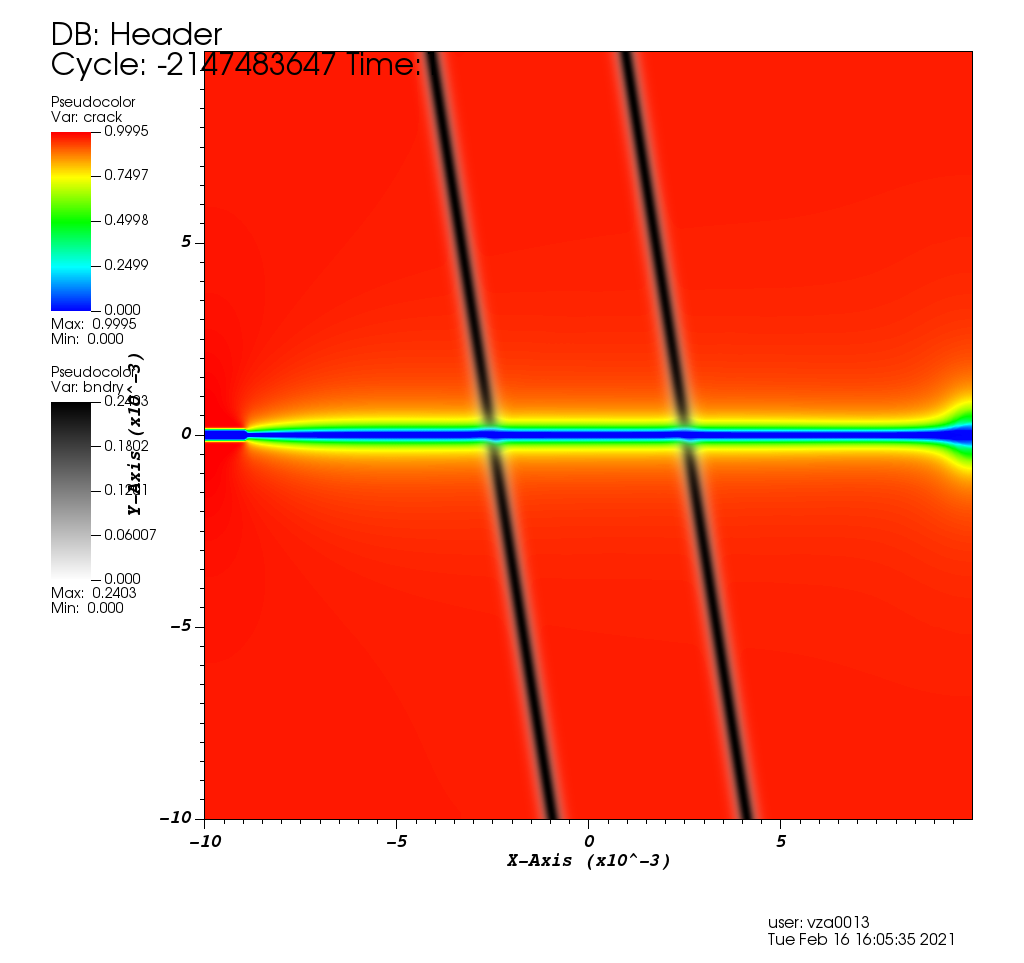} \caption{}\label{fig:laminate_output_1734}\end{subfigure}
  \end{minipage}
  \end{minipage}

  \caption{Laminate: Visualization of crack field corresponding to cases A-F highlighted in Figure \ref{fig:angled_interface} right. 
  The six representative cases demonstrate {\color{C0}\bf pure fracture}, {\color{C2}\bf delamination} \& {\color{C1}\bf transition} behavior of the crack.
  The laminate interface is overlaid in black for visualization purposes.}
  \label{fig:planar_laminate_visualizations}
\end{figure}

\begin{figure}
  \centering 
  \includegraphics[width=\linewidth]{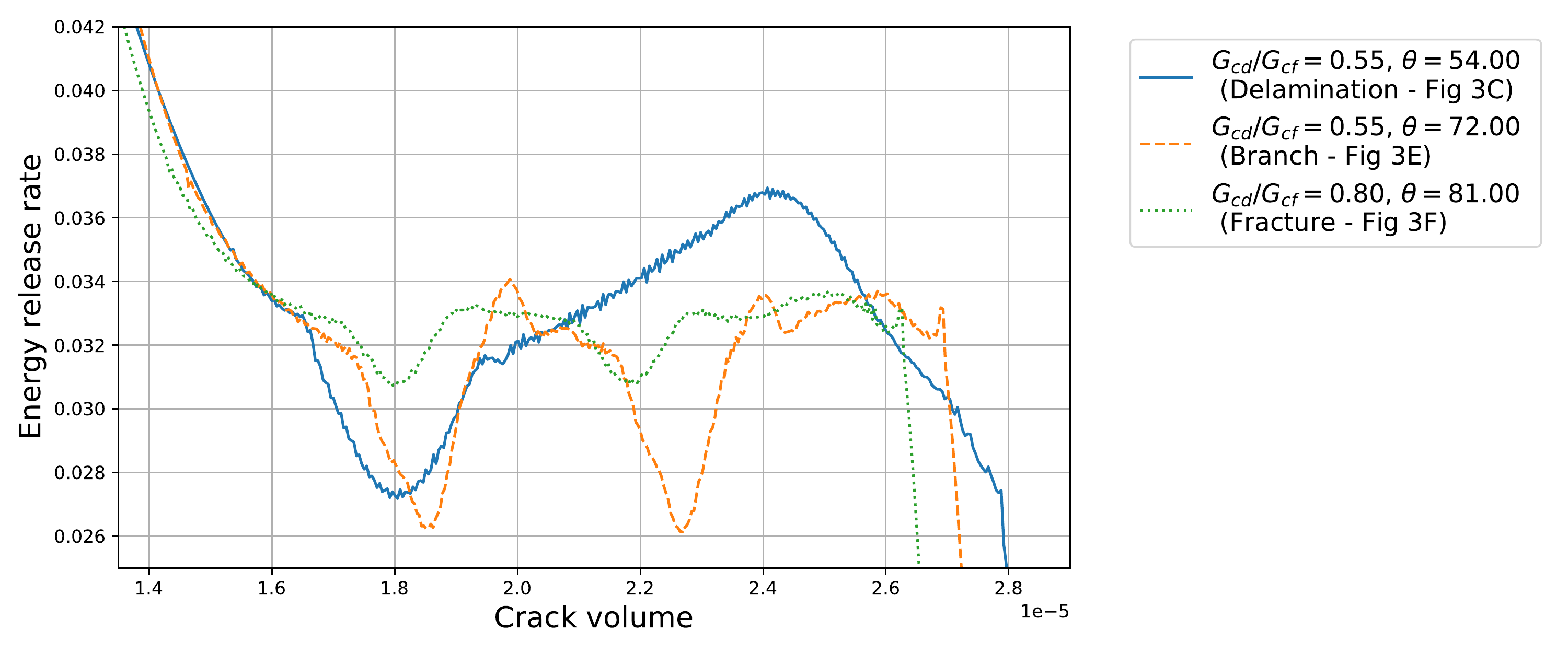}
  \caption{Laminate: Plot depicting the evolution of fracture energy release rate as crack propagates through the domain. 
  }
  \label{fig:Laminate_ERR}
\end{figure}

In this set of simulations we consider a crack impinging an angled laminate inclusion.
The interface has a delamination energy release rate $G_{cd}$, which is varied from $0.5G_{cf}$ to $G_{cf}$.
We vary the orientation of the laminate inclusion from $0^\circ$ (vertical) to $90^\circ$ (horizontal) (Figure~\ref{fig:angled_interface} Left).
We investigate the relationship between the non-dimensional parameters $G_{cd}/G_{cf}$ and the orientation $\theta$ to determine the regions in which delamination and fracture are dominant.

In the case of ideal crack propagation, it is possible to estimate these two regions using a simple analytic calculation.
If the crack tip has just encountered the interface, the transition occurs when the differential energy released by fracture $dE_f$ equals that of delamination $dE_d$.
The energies are related to the crack length as
\begin{align}
    dE_f &\sim G_{cf}da_f &  dE_d &\sim G_{cd} da_d
\end{align}
where $da_f$ and $da_d$ are the differential area increases corresponding to fracture and delamination, respectively.
Since relationship between the areas is $da_f = da_d\sin\theta,$ the transition between delamination and fracture clearly occurs when
\begin{align}\label{eq:laminate_analytic}
    \frac{G_{cd}}{G_{cf}} = \sin\theta.
\end{align}
We emphasize that this relation does not account for the additional energy required to redirect the crack in the event of delamination, which is apparently nontrivial.
Therefore, we expect delamination to be somewhat less likely than predicted by (\ref{eq:laminate_analytic}).

A total of 110 simulations were performed with varying angle $\theta$ and $G_{cd}/G_{cf}$ ratio (Figure~\ref{fig:angled_interface}).
Each simulation is colored based on the observed behavior, with green corresponding to delamination and blue to fracture.
Delamination generally correspond to a clean crack propagation along the interface (Figure~\ref{fig:laminate_output_1035}).
However, it was observed when the laminate tilt angle exceeded $45^\circ$, that the delamination cracks generally redirected into the material as a fracture (Figure~\ref{fig:laminate_output_1776}).
This is a result of edge effects, and these events are still categorized as pure delamination.
Pure fracture events were identified as cracks that were generally unaffected by the presence of the interface (Figure~\ref{fig:laminate_output_1790}) or where the effect of the interface was minimal (Figure~\ref{fig:laminate_output_1734}).
The pure fracture behavior was generally identical to classic Mode I fracture.

There were a substantial number of ``transition'' results, which are all indicated in yellow.
These are cracks that were neither purely delamination nor purely fracture, but contained some combination of both.
Some cracks delaminated for a very short period but then split off to fracture the material (Figure~\ref{fig:laminate_output_1037}).
Crack branching was very common, in which the mode was predominantly fracture combined with small delaminations at the crack-interface intersection.
There was a continuum between transition and fracture, and the categorization of a pattern as transition vs fracture is somewhat subjective.
The primary bifurcation occurs at the line between the transition and the delamination points.
The match between the observed fracture/delamination bifurcation point and the highly simplified, idealized analytic solution appears to be reasonable.

Figure \ref{fig:Laminate_ERR} shows the evolution of the fracture energy release rate $G$ as the crack moves through the domain for different cases.
In the case of delamination ($G_{cd}/G_{cf} = 0.55$, $\theta = 54^\circ$, Figure \ref{fig:planar_laminate_visualizations}c), the energy release rate first shows a sharp dip followed by an increase and then finally a dip. 
The first dip corresponds to when the crack first encounters an interface resulting in crack redirection and delamination.
As the crack progresses traveling in an energetically low trajectory, the stress increases and so does $G$. 
The crack finally reflects off the boundary into a fracture mode of failure resulting in a decline in $G$.
In the case of fracture ($G_{cd}/G_{cf} = 0.55$, $\theta = 54^\circ$, Figure \ref{fig:planar_laminate_visualizations}f), the crack stays mostly constant except for two dips corresponding to the point is encounters the two interfaces.
Finally, for the case of branching ($G_{cd}/G_{cf} = 0.55$, $\theta = 72^\circ$, Figure \ref{fig:planar_laminate_visualizations}e), the energy release rate shows bigger and wider dips.
After encountering the first interface, the energy release rate dips and the crack splits off.
Due to higher energies associated with crack deflection, the delaminated section of the crack eventually stops.
This causes the energy release rate to stabilize.
The process repeats itself when the second interface is encountered.

\subsection{Wavy interface}

\begin{figure}
  \begin{minipage}[t][6cm][t]{\linewidth}
    \begin{minipage}[t][6cm][t]{0.3\linewidth}
      \includegraphics[width=\linewidth,clip,trim=0cm -1.3cm 0cm 0cm]{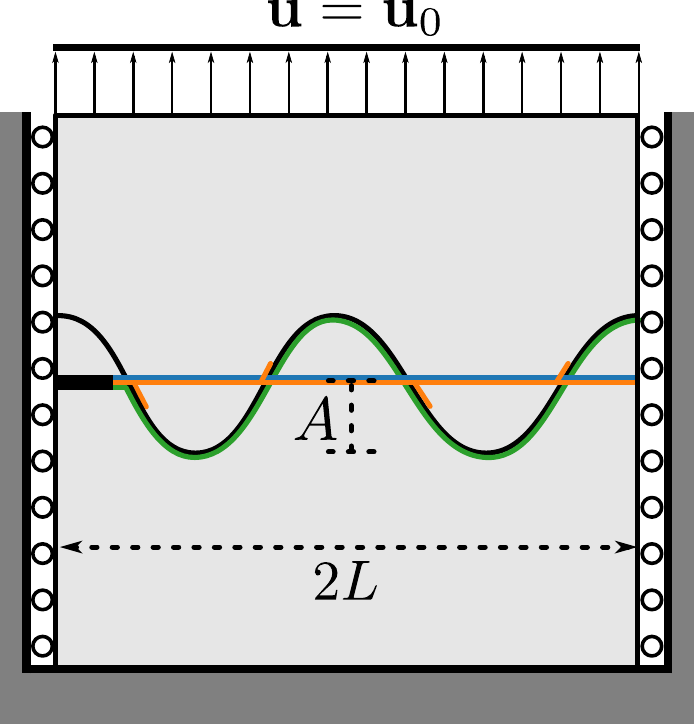}
    \end{minipage}\hfill
    \begin{minipage}[t][6cm][t]{0.6\linewidth}
      \includegraphics[width=\linewidth]{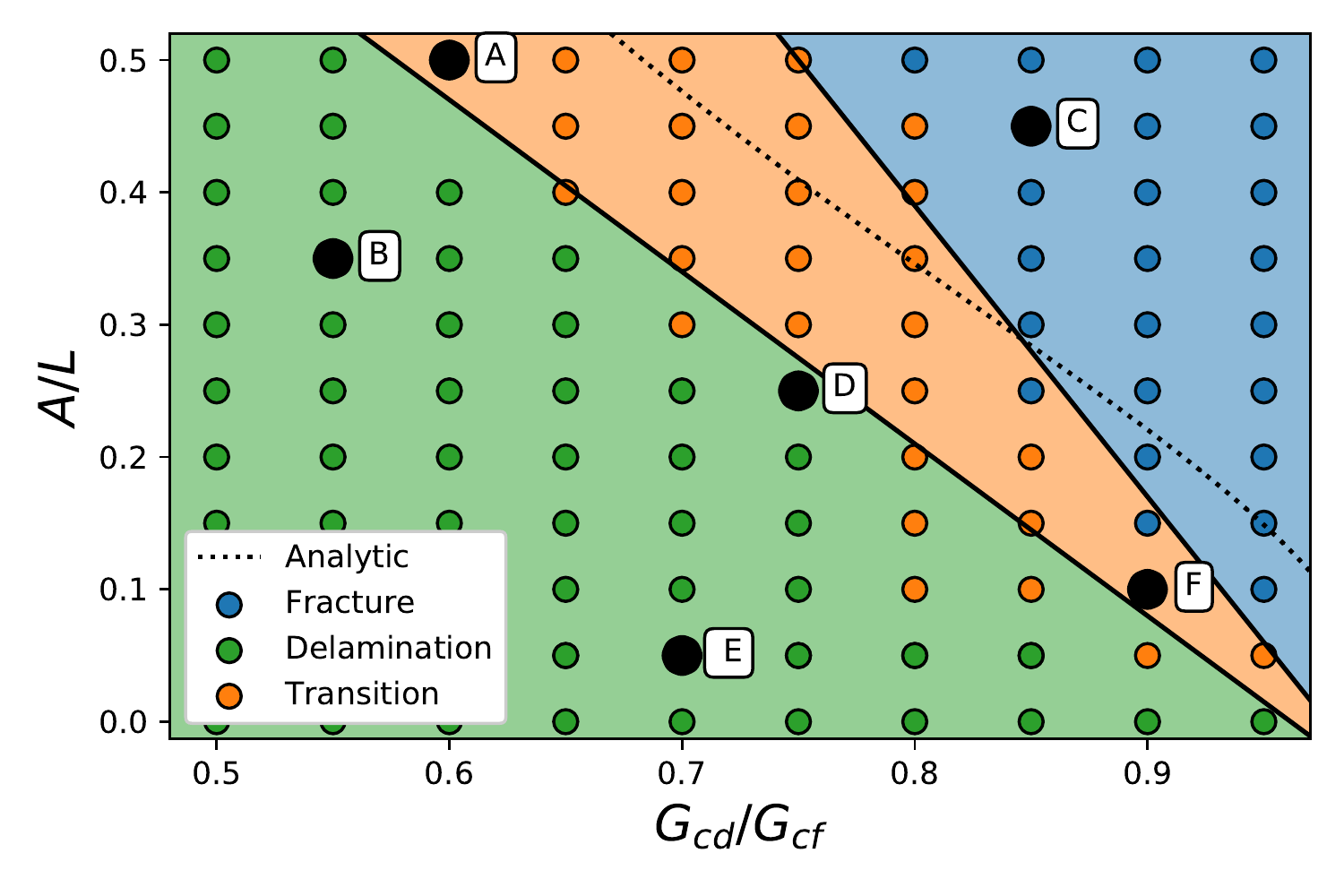}
    \end{minipage}\hfill
  \end{minipage}
  \caption{Wavy interface: ({\bf Left}) Schematic of the sinusoidal interface problem. 
    Colored lines indicate sample crack paths for ease of visualization.
    Colors correspond to delamination (green), transition (orange) and fracture (blue).
    ({\bf Right}) Paramater survey of delamination/fracture for a sinusoidal interface identifying delamination, fracture and transition zones. 
    Callouts A-F correspond to visualization of crack paths presented in 
    Figure \ref{fig:VisualizationWavyInterface}.
    Simplified analytic prediction is overlaid for comparison.  
    (Compare: \cite{sehr2019interface} Figure 10)
  }
  \label{fig:wavy_interface}
\end{figure}

\begin{figure}[h]
  \centering
  \begin{minipage}{0.15\linewidth}
    \includegraphics{figures/legend_vertical.png}
  \end{minipage}
  \begin{minipage}{0.75\linewidth}
    \begin{minipage}{\linewidth}
      \begin{subfigure}{0.3\linewidth}\includegraphics[width=\linewidth,clip,trim=7.2cm 5.5cm 1.7cm 3.5cm,cfbox=C1 3pt 0pt]{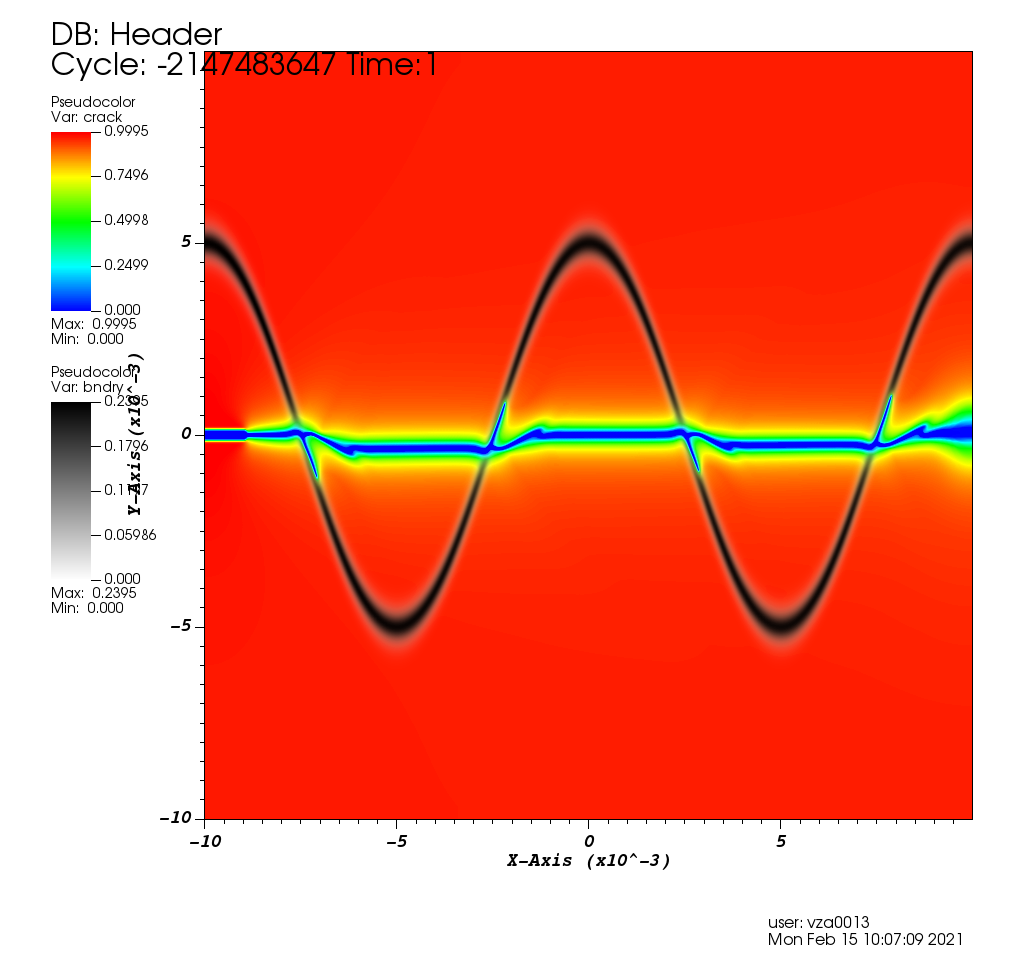} \caption{}\end{subfigure}\hspace{10pt}%
      \begin{subfigure}{0.3\linewidth}\includegraphics[width=\linewidth,clip,trim=7.2cm 5.5cm 1.7cm 3.5cm,cfbox=C2 3pt 0pt]{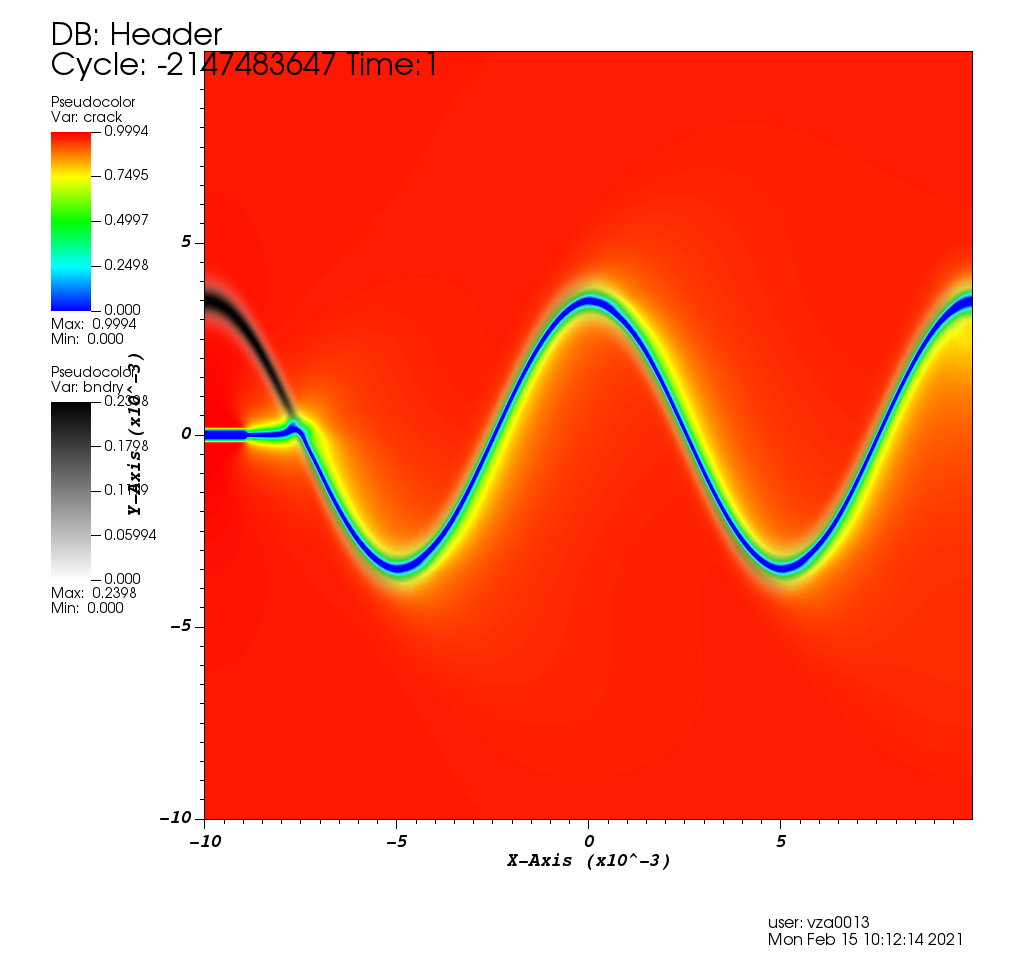} \caption{}\end{subfigure}\hspace{10pt}%
      \begin{subfigure}{0.3\linewidth}\includegraphics[width=\linewidth,clip,trim=7.2cm 5.5cm 1.7cm 3.5cm,cfbox=C0 3pt 0pt]{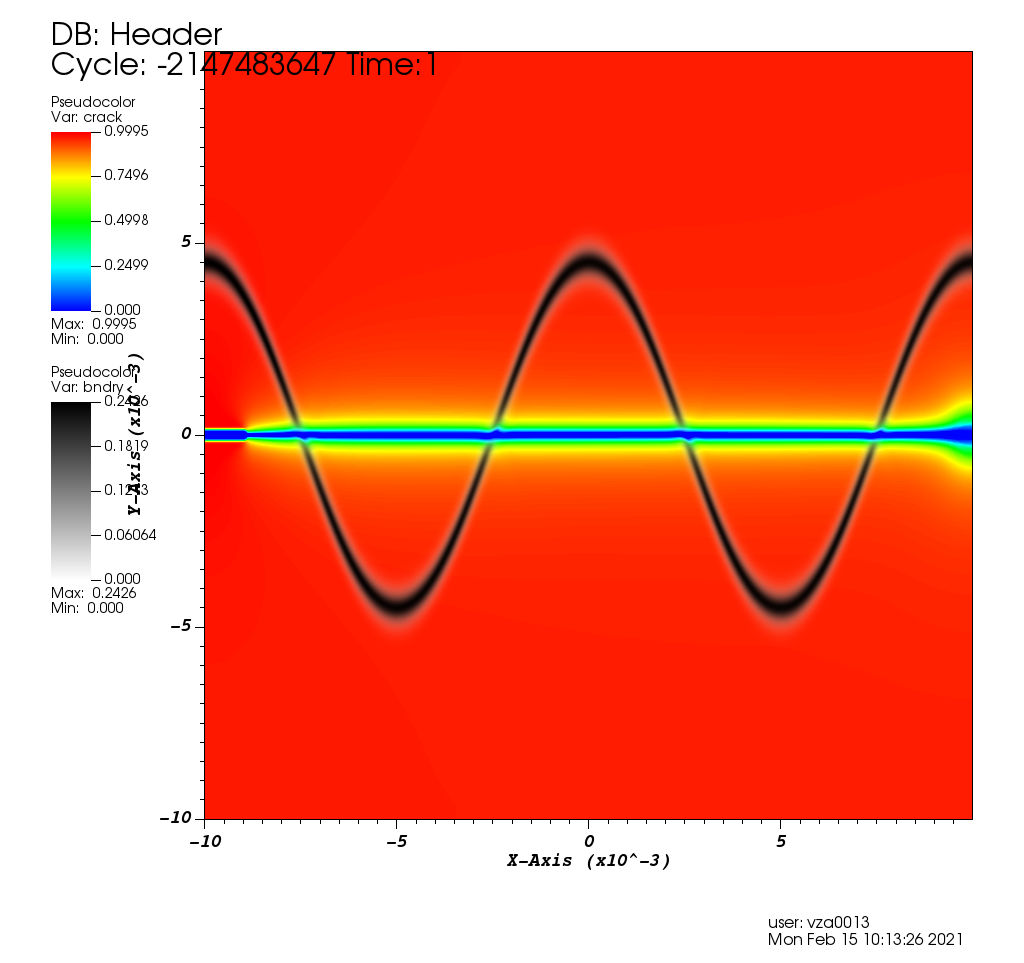} \caption{}\end{subfigure}%
    \end{minipage}
    \begin{minipage}{\linewidth}
      \begin{subfigure}{0.3\linewidth}\includegraphics[width=\linewidth,clip,trim=7.2cm 5.5cm 1.7cm 3.5cm,cfbox=C1 3pt 0pt]{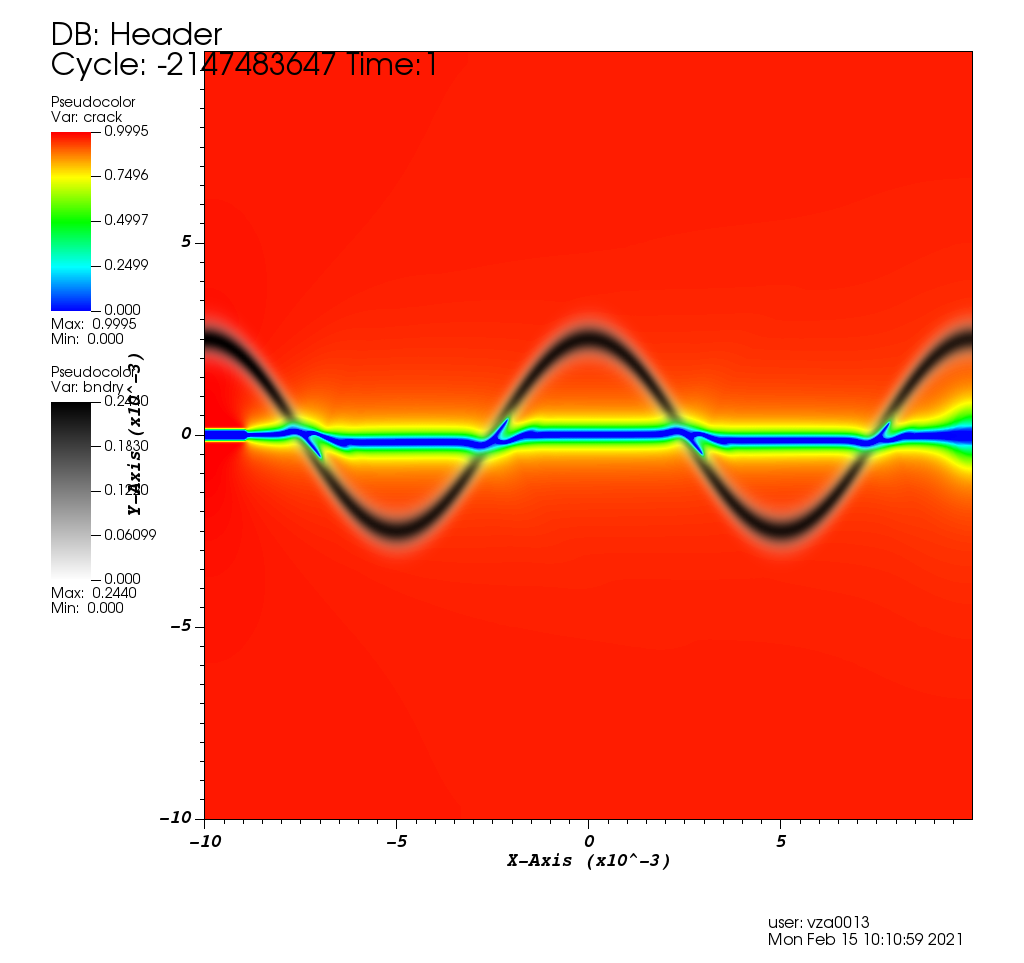} \caption{}\end{subfigure}\hspace{10pt}%
      \begin{subfigure}{0.3\linewidth}\includegraphics[width=\linewidth,clip,trim=7.2cm 5.5cm 1.7cm 3.5cm,cfbox=C2 3pt 0pt]{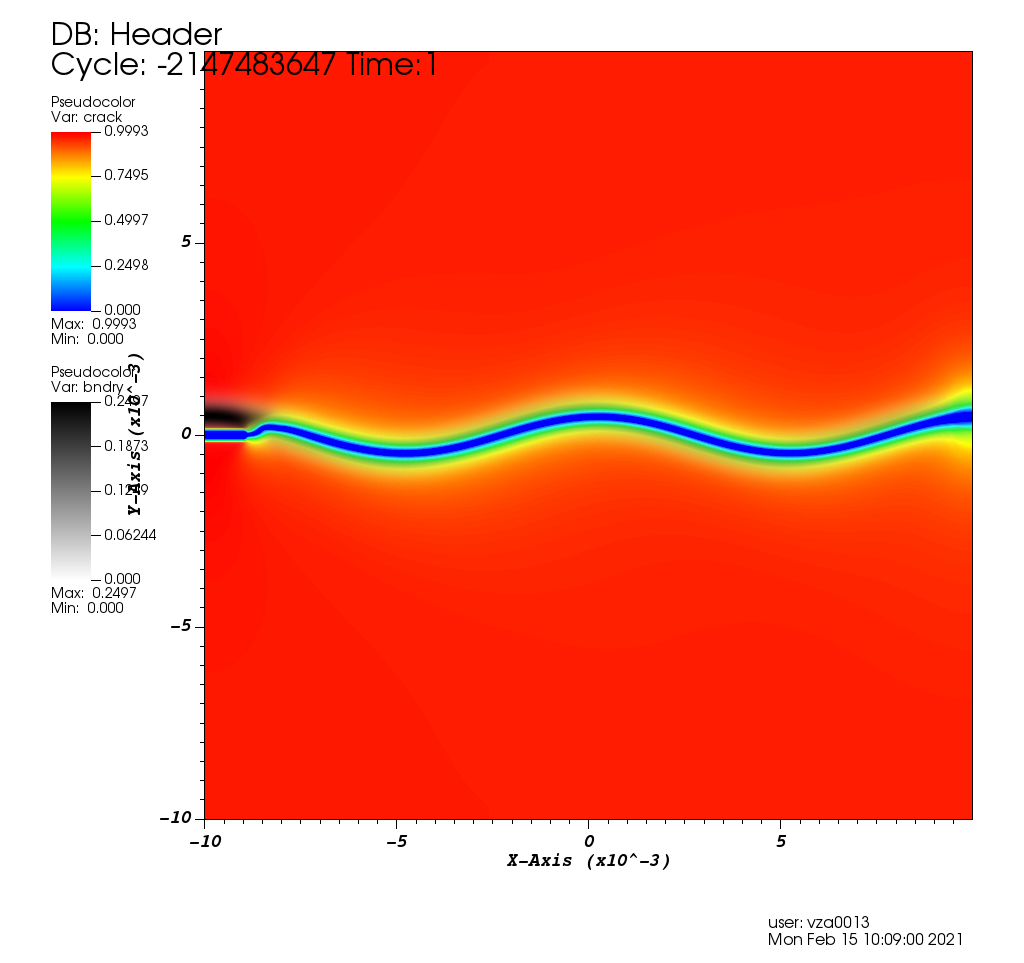} \caption{}\end{subfigure}\hspace{10pt}%
      \begin{subfigure}{0.3\linewidth}\includegraphics[width=\linewidth,clip,trim=7.2cm 5.5cm 1.7cm 3.5cm,cfbox=C0 3pt 0pt]{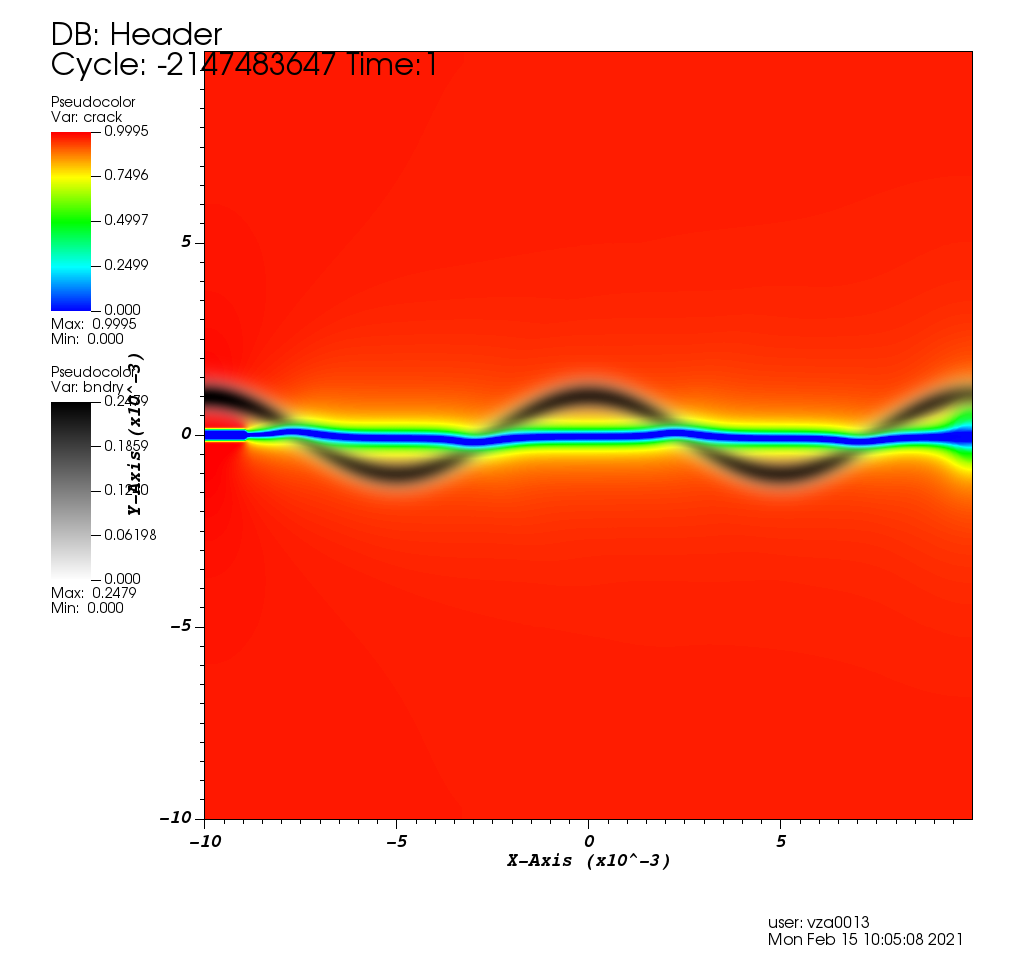} \caption{}\end{subfigure}
    \end{minipage}
  \end{minipage}

  \caption{Wavy interface: Visualization of a selection of crack behavior. The wavy interface is marked in gray. Each figure corresponds to a point on Figure \ref{fig:wavy_interface}. The outline boxes denote {\bf{\color{C1}branching}}, {\bf{\color{C2}delamination}}, and {\bf{\color{C0}fracture}}}
  \label{fig:VisualizationWavyInterface}
\end{figure}

\begin{figure}[h]
  \centering 
  \includegraphics[width=\linewidth]{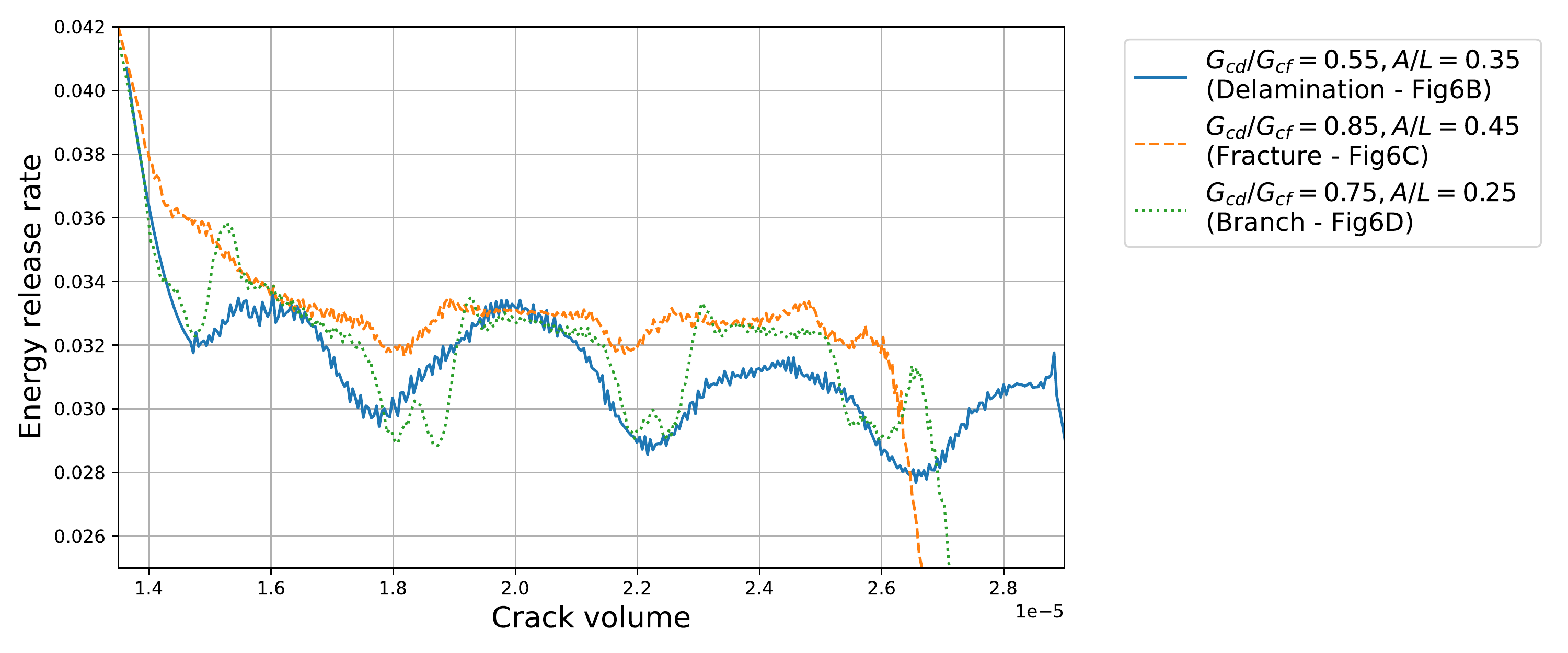}
  \caption{Wavy Interface: Plot depicting the evolution of fracture energy release rate as crack propagates through the domain. 
  }
  \label{fig:Wavy_ERR}
\end{figure}

Delamination vs fracture behavior in non-planar interfaces is of great importance in any application involving joined materials.
Classic examples include material systems produced by thermal spraying \cite{hsueh1999effects}, textile ceramic composites \cite{blacklock2016virtual}, and bio-inspired sutured interfaces \cite{li2013generalized, liu2020interfacial,cao2019experimental}
In this section we consider the behavior of a crack under Mode I loading along a sinusoidally varying interface with $G_{cd}\le G_{cf}$.
Once again, the interface was diffused with a length scale $b=5\times 10^{-4}$.
In such a configuration, there is an obvious competition between the lower energetic cost of delamination and the lower amount of surface area resulting from fracture.
As with the previous example there are three anticipated possibilities: pure fracture, pure delamination, and transitionary behavior comprising elements of both (Figure~\ref{fig:wavy_interface} Left).

Following the above, we can estimate the transition point between delamination and fracture via a simple by-hand calculation.
The relationship between the fracture and delamination areas is slightly more complex.
Assuming a homogenized interface and that the crack propagates uniformly, then the area elements are related by the arc length formula as
\begin{equation}
    da_f 
    = \Bigg(\frac{2}{\pi} \int_0^{\pi/2}\sqrt{1 + \frac{\pi^2 A^2}{L^2}\sin^2(x)}\,dx\Bigg) da_d
    = \frac{2}{\pi} E\Big(-\frac{\pi^2 A^2}{L^2}\Big) \,da_d,
\end{equation}
where $E$ is the compete elliptic integral of the second kind, $A$ is the amplitude of the sinudoid, and $L$ is the period.
Therefore, we find the relationship between the $Gc$ ratio and the ratio $A/L$ to be 
\begin{align}
    \frac{G_{cd}}{G_{cf}} = \frac{\pi}{2E(-\pi^2A^2/L^2)}.
\end{align}
We emphasize that this highly simplified calculation does not account for curvature effects, for mode II effects, or for the energetic cost associated with crack kinking.
Therefore we expect this transition point to be, in many cases, artificially high.

We considered 110 configurations with $A/L$ ranging from $0$ to $0.5$ and $G_{cd}/G_{cf}$ ratios from 0.5 to 1.0.
As before, we identify three zones: {\color{C0}\bf pure fracture}, {\color{C2}\bf delamination}, and {\color{C1}\bf transition} (Figure~\ref{fig:wavy_interface} right). 
As before, there were significant transitionary cases where the crack starts to delaminate, but then splits off to fracture the material. 
As expected the actual transition happens much lower than the transition point obtained from the analytical predictions.
Figure \ref{fig:VisualizationWavyInterface} shows actual crack path for six different cases corresponding to separate zones marked in Figure \ref{fig:wavy_interface} demonstrating transition from branching to delamination to fracture for different amplitudes and $G_{cd}/G_{cf}$ ratios.

Figure \ref{fig:Wavy_ERR} shows the evolution of fracture energy release rate $G$ as the crack propagates leading to delamination, fracture and branching.
As expected, for pure fracture, $G$ stabilizes and remains constant for most of the crack path, and finally dropping off.
The final drop off in $G$ happens because the crack reaches the edge of the domain.
For delamination, the $G$ shows wave-like behavior as the crack follows the interface leading to delamination.
For branching, the $G$ curve exhibits fluctuations upon encountering the interface, and then proceeds with a constant value.
These results are consistent with \cite{sehr2019interface} who studied kinking vs delamination behavior for cracks using cohesive zone modeling with different amplitude $G_{cd}/G_{cf}$ ratios.

\subsection{Circular inclusion}
Crack deflection and redirection in heterogeneous materials has been a topic studied through experimental and numerical methods.
In this section, we consider an idealized problem of a linear elastic isotropic circular inclusion of radius $R$ and modulus $E_i$ embedded in a linear elastic isotropic matrix of modulus $E_m$.
We choose the same fracture energy $G_{cf}$ of both materials but vary the delamination fracture energy of the interface $G_{cd}$.
Finally, we vary the offset $h$ of the center of the inclusion from the geometric center of the square domain.
We study the role of non-dimensional parameters $E_i/E_m$, $G_{cd}/G_{cf}$ and $h/R$ on potential crack paths.
As with the previous two cases, crack behavior may be categorized as:) penetration of the inclusion ({\color{C0}\bf inclusion fracture}), delamination of the interface ({\color{C2}\bf delamination}) or a combination ({\color{C1}\bf branching}).
Here, an additional possibility is that the crack does not interact with the inclusion at all, perhaps as a result of deflection ({\color{C3}\bf matrix fracture}).

\begin{figure}
  \begin{minipage}[t][6cm][t]{\linewidth}
    \begin{minipage}[t][6cm][t]{0.3\linewidth}
      \includegraphics[width=\linewidth,clip,trim=0cm -1.3cm 0cm 0cm]{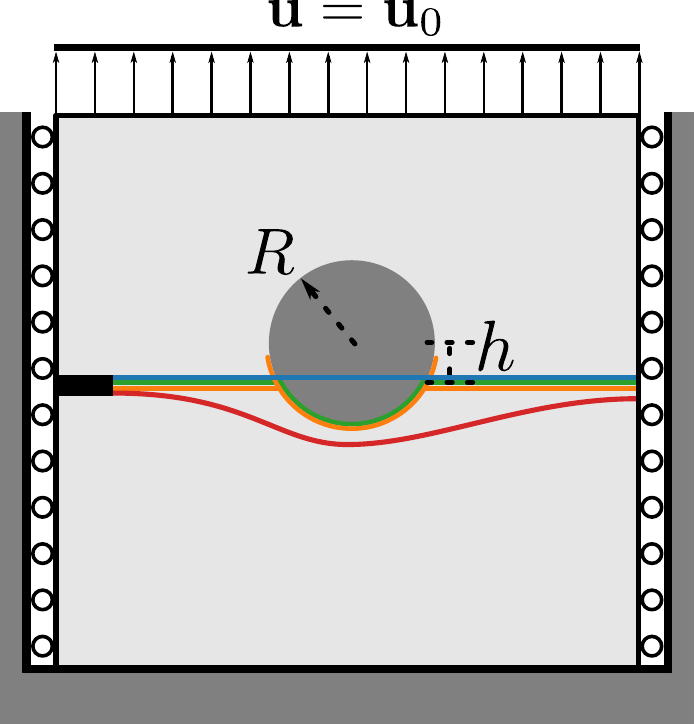}
    \end{minipage}\hfill
    \begin{minipage}[t][6cm][t]{0.6\linewidth}
      \includegraphics[width=\linewidth]{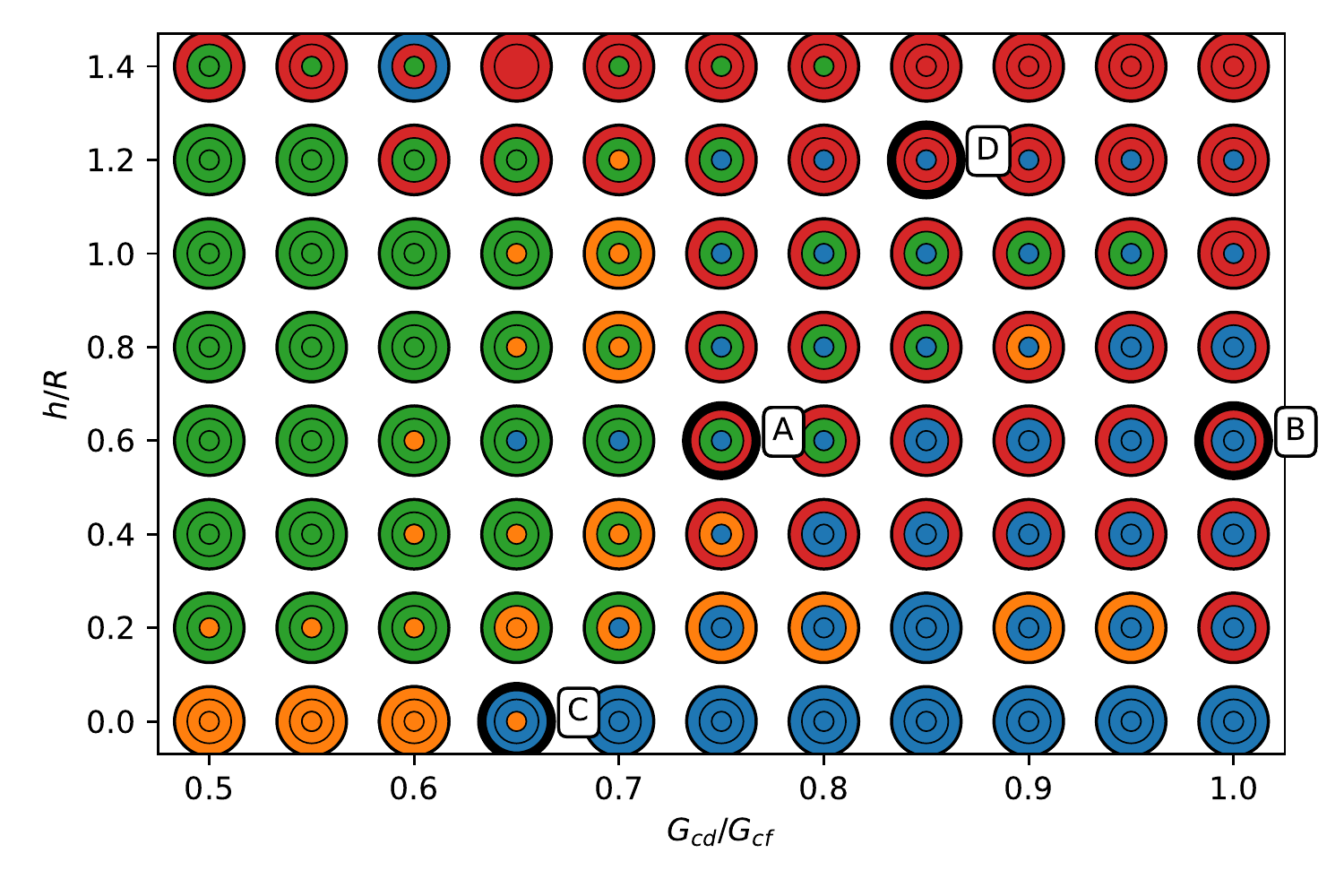}
    \end{minipage}\hfill
  \end{minipage}
  \caption{
    Inclusion: 
    ({\bf Left}) Schematic of the problem and observed crack paths. ({\bf Right}) Plot depicting different outcomes for different scenarios. Each point corresponds to a specific $G_{cd}/G_{cf}$ ratio and offset. The inner circles correspond to a modulus ratio $E_{i}/E_{m}$ of $0.5$, the middle circles to a modulus ratio of $1$, and the outer circles to a modulus ratio of $2.0$. Colors correspond to {\color{C3}\bf matrix fracture}, {\color{C0}\bf pure inclusion fracture}, {\color{C2}\bf delamination}, and {\color{C1}\bf branching}.}
  \label{fig:crack_deflection}
\end{figure}

Figure \ref{fig:crack_deflection} presents a summary plot for different crack behavior for varying offset ratios $h/R$, fracture energy ratios $G_{cd}/G_{cf}$, and modulus ratios $E_i/E_m$. 
As expected, the observed behavior is highly complex, and often accompanied by transition modes. 
In general, for low $G_{cd}/G_{cf}$ values, delamination (depicted by green) is the predominant mode of failure even for high offset ratios.
For high offset ratios, the crack path transitions from delamination to pure matrix fracture as $G_{cd}/G_{cf}$ increases. 
For zero offset ratio, the crack path transitions from a branching behavior to pure inclusion fracture as $G_{cd}/G_{cf}$ increases. 
This is to be expected since the energetic cost of delamination increases as $G_{cd}/G_{cf}$ ratio increases.
For a fixed value of $G_{cd}/G_{cf}$ ratio, the crack path changes from pure inclusion fracture to pure matrix fracture as offset  ratio increases.
Once again, this is to be expected because as the offset ratio increases, the cost of crack deflection also increases. 

\begin{figure}
  \begin{minipage}{\linewidth}
    \centering \includegraphics{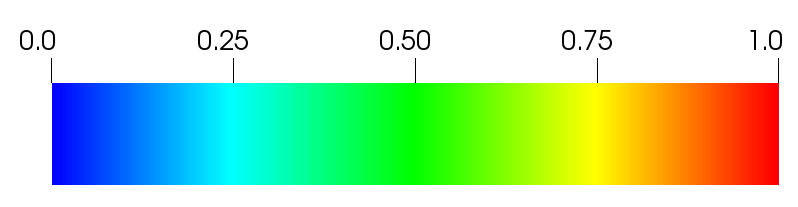}
  \end{minipage}

  \begin{minipage}{0.05\linewidth}
    \rotatebox{90}{$E_{i}/E_{m}=0.5$}
  \end{minipage}
  \begin{minipage}{0.95\linewidth}
    \includegraphics[width=0.23\linewidth,clip,trim=7.2cm 5.5cm 1.7cm 3.5cm,cfbox=C0 3pt 0pt]{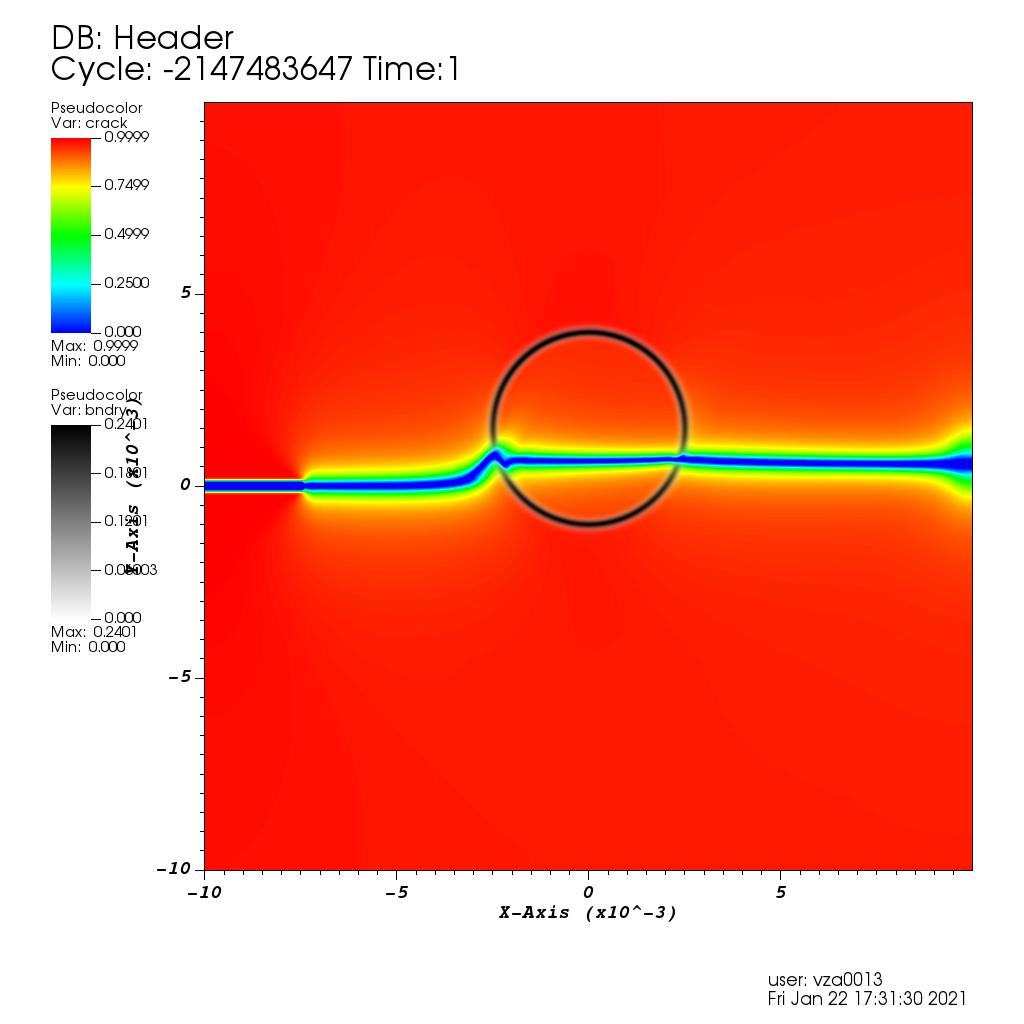}\hfill
    \includegraphics[width=0.23\linewidth,clip,trim=7.2cm 5.5cm 1.7cm 3.5cm,cfbox=C0 3pt 0pt]{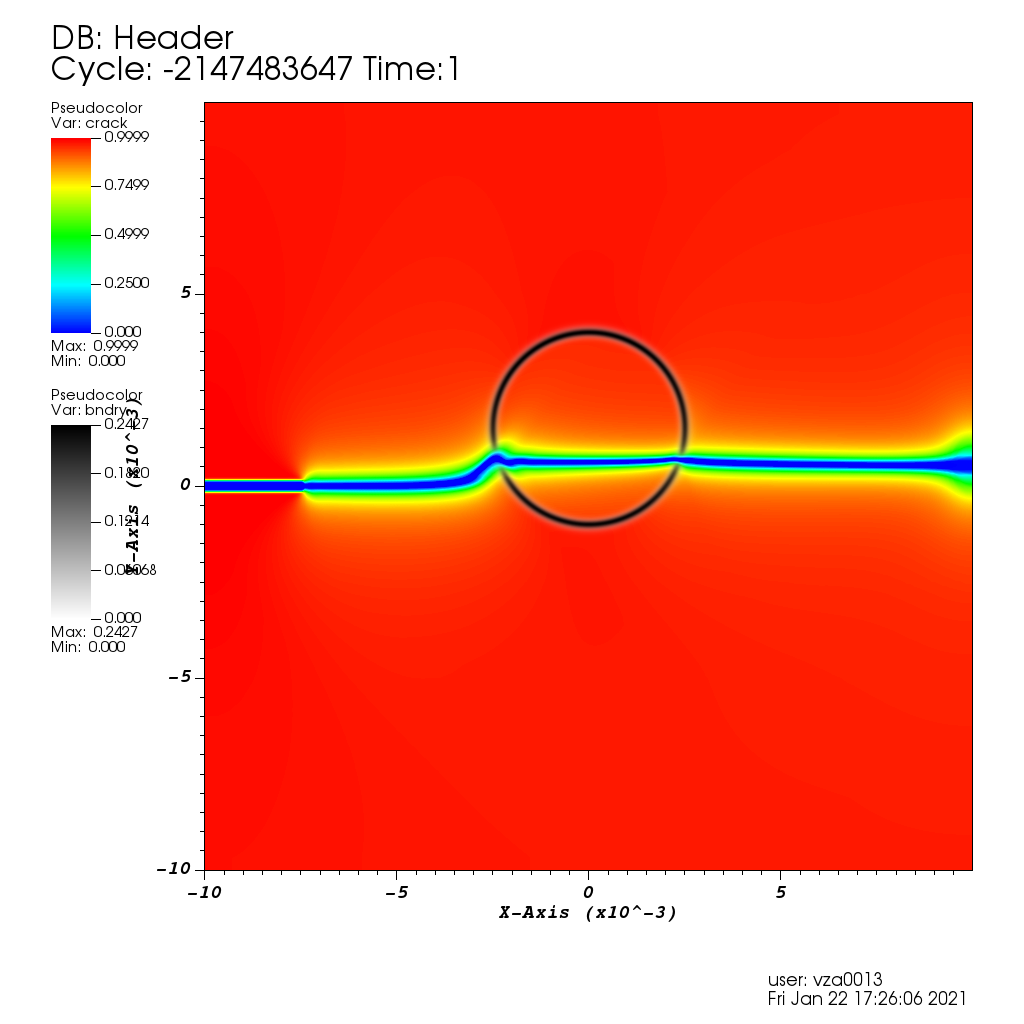}\hfill
    \includegraphics[width=0.23\linewidth,clip,trim=7.2cm 5.5cm 1.7cm 3.5cm,cfbox=C1 3pt 0pt]{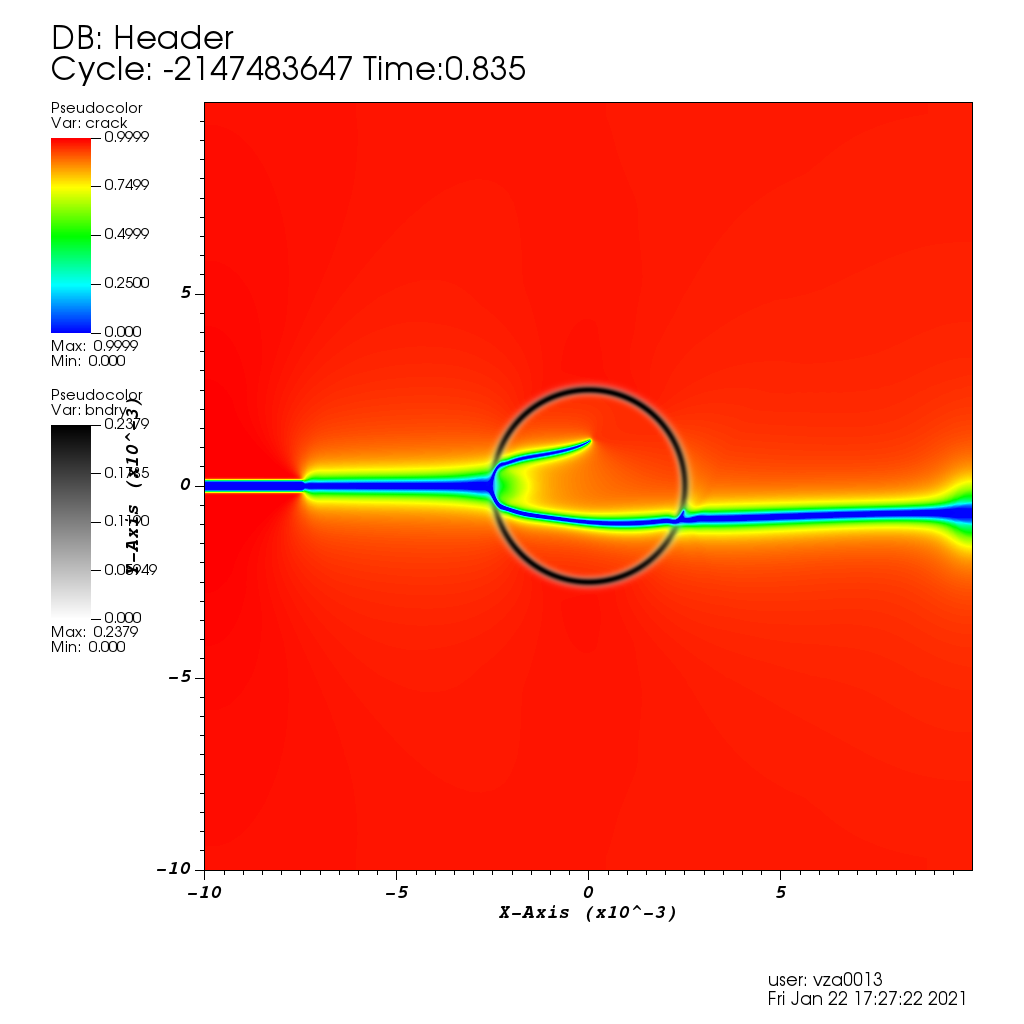}\hfill
    \includegraphics[width=0.23\linewidth,clip,trim=7.2cm 5.5cm 1.7cm 3.5cm,cfbox=C0 3pt 0pt]{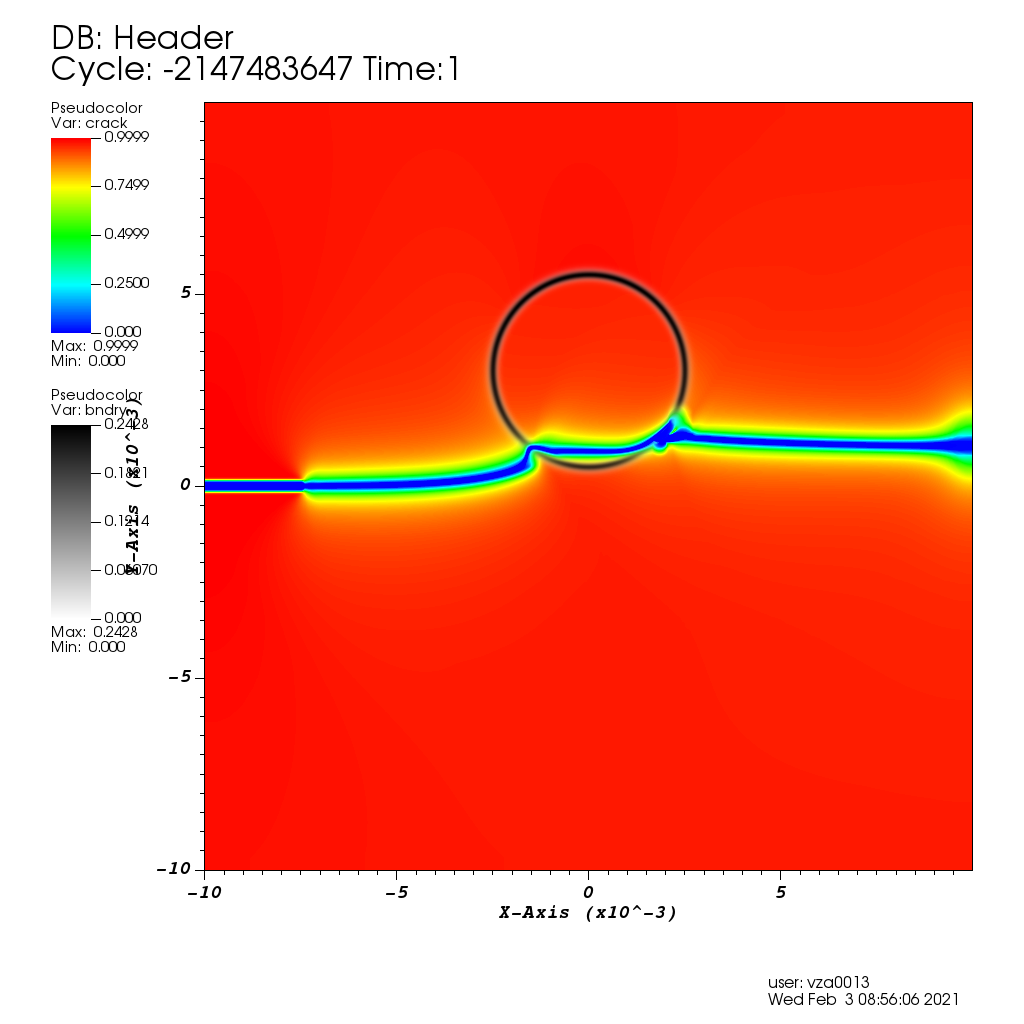}
  \end{minipage}
  \begin{minipage}{0.05\linewidth}
    \rotatebox{90}{$E_{i}/E_{m}=1.0$}
  \end{minipage}
  \begin{minipage}{0.95\linewidth}
    \includegraphics[width=0.23\linewidth,clip,trim=7.2cm 5.5cm 1.7cm 3.5cm,cfbox=C2 3pt 0pt]{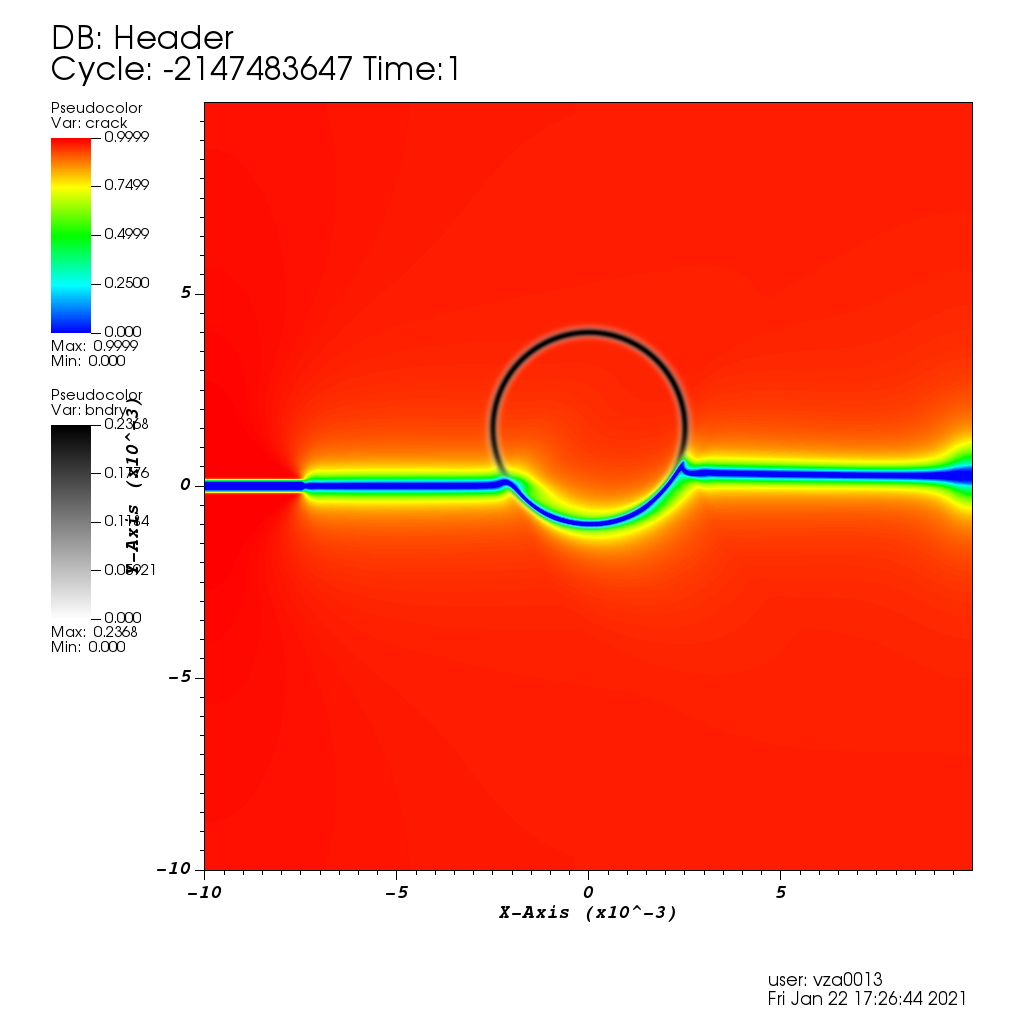}\hfill
    \includegraphics[width=0.23\linewidth,clip,trim=7.2cm 5.5cm 1.7cm 3.5cm,cfbox=C0 3pt 0pt]{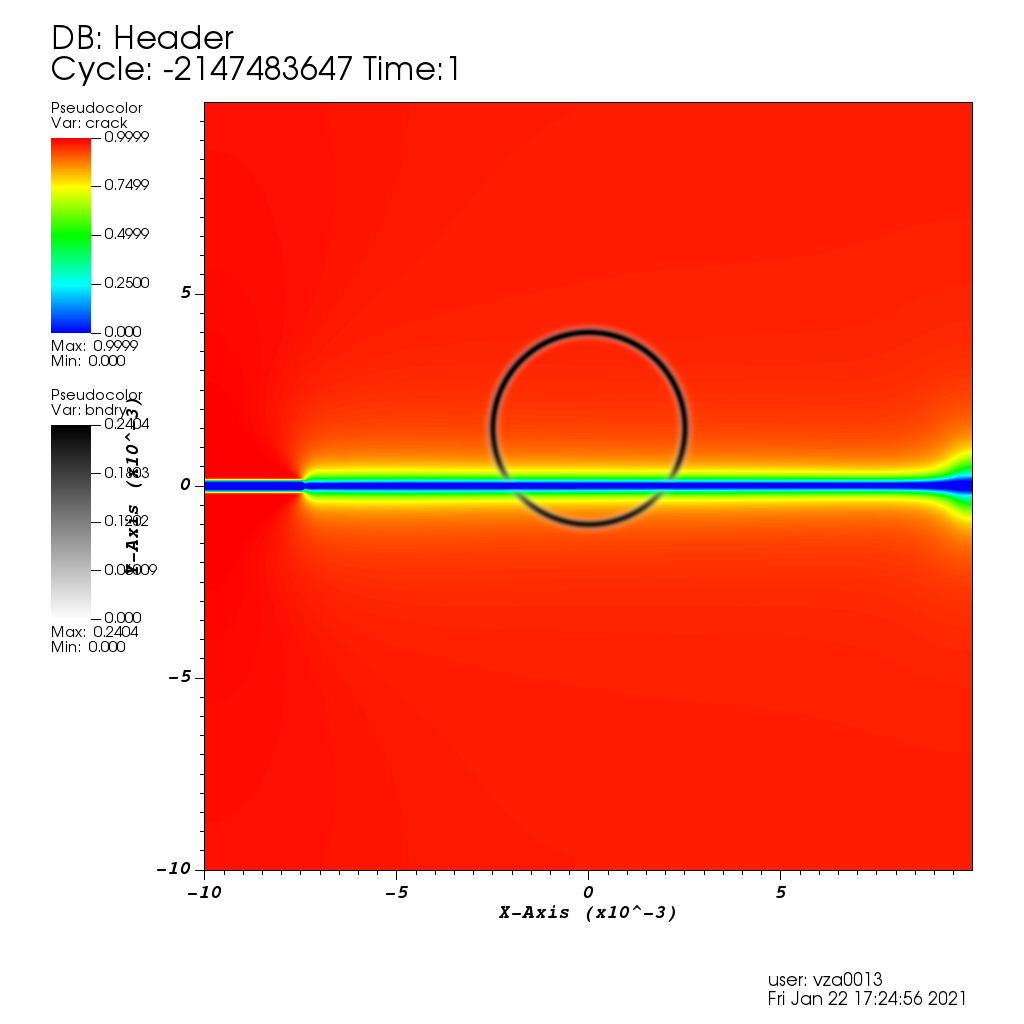}\hfill
    \includegraphics[width=0.23\linewidth,clip,trim=7.2cm 5.5cm 1.7cm 3.5cm,cfbox=C0 3pt 0pt]{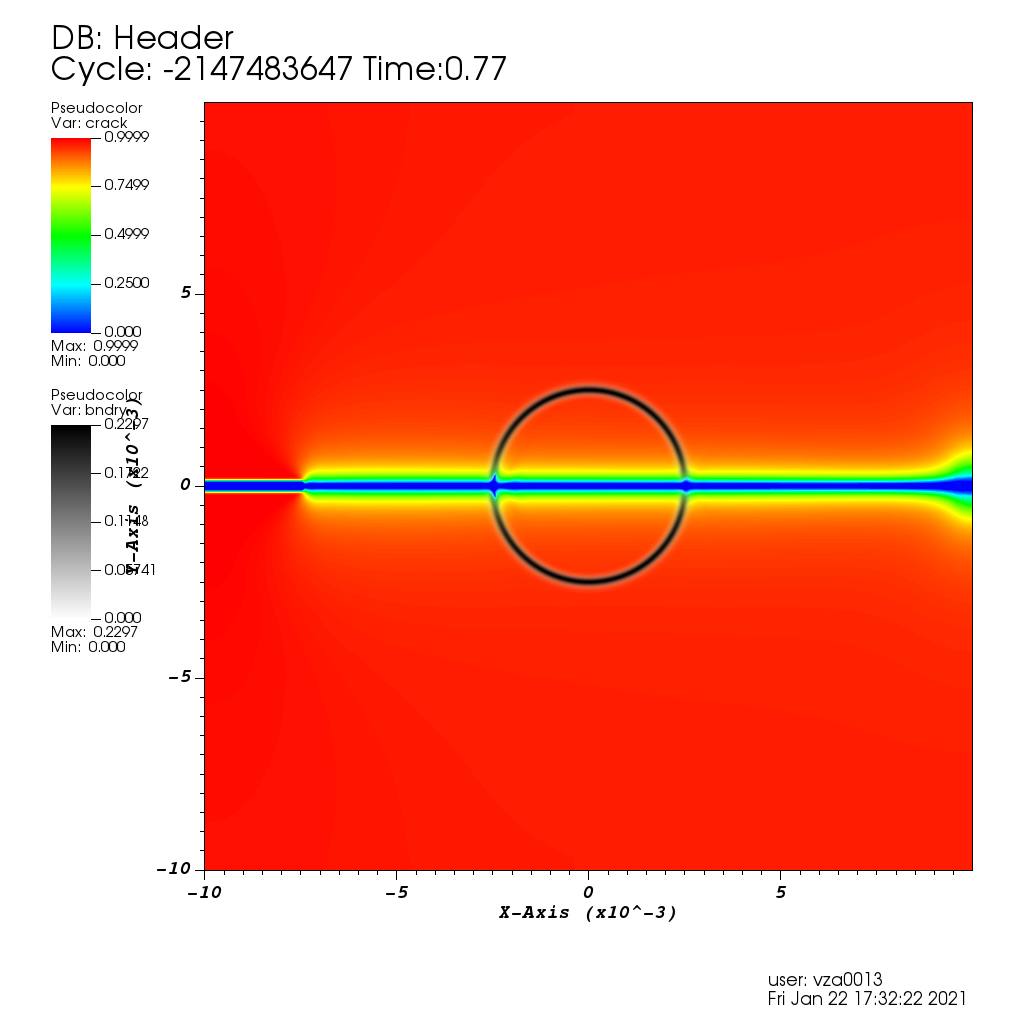}\hfill
    \includegraphics[width=0.23\linewidth,clip,trim=7.2cm 5.5cm 1.7cm 3.5cm,cfbox=C3 3pt 0pt]{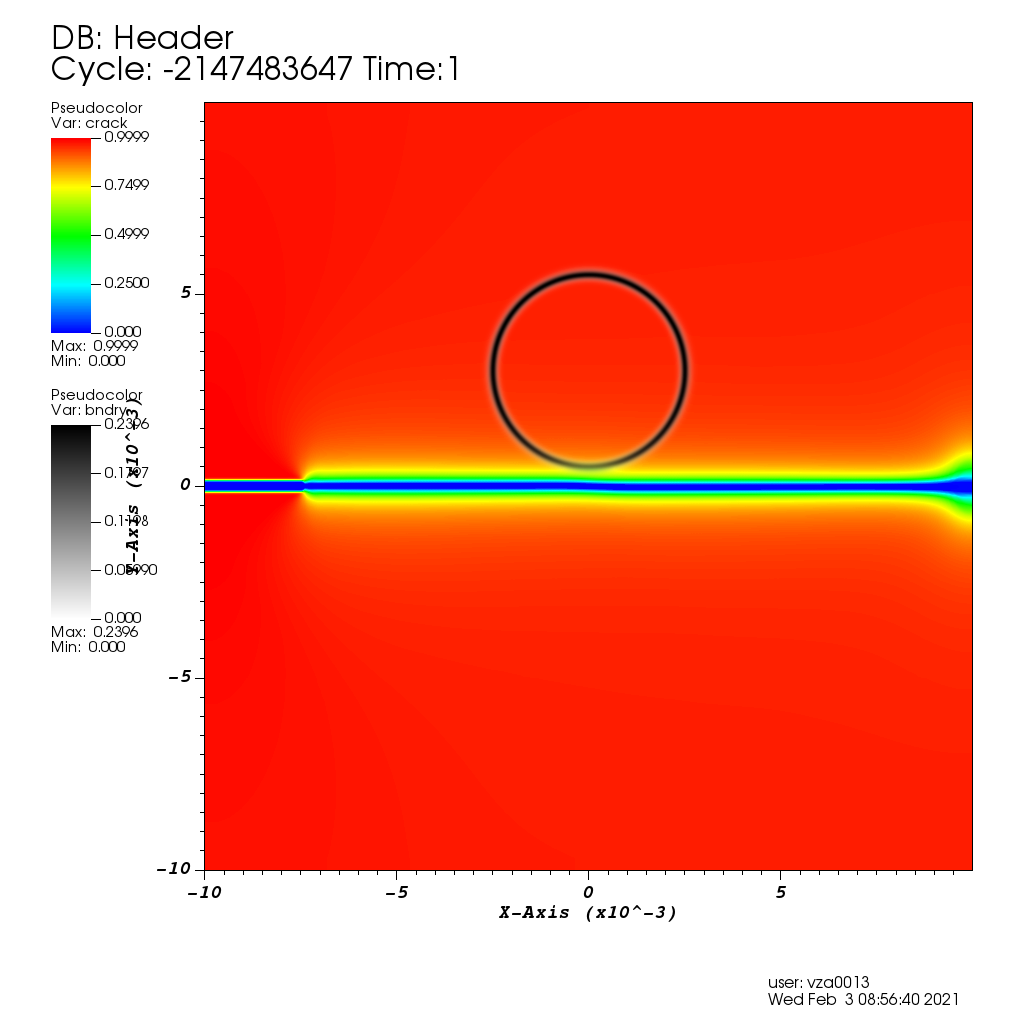}
  \end{minipage}
  \begin{minipage}{0.05\linewidth}
    \rotatebox{90}{$E_{i}/E_{m}=2.0$}
  \end{minipage}
  \begin{minipage}{0.95\linewidth}
    \includegraphics[width=0.23\linewidth,clip,trim=7.2cm 5.5cm 1.7cm 3.5cm,cfbox=C3 3pt 0pt]{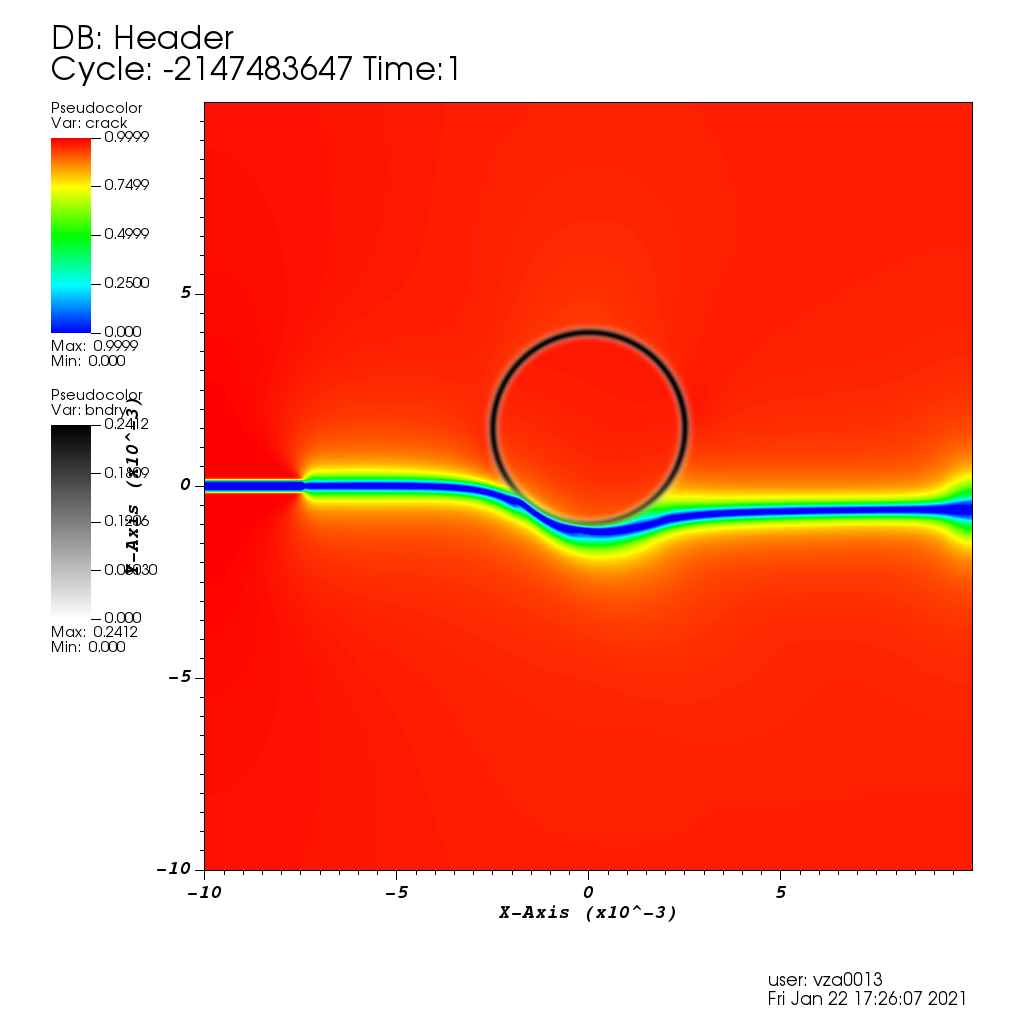}
    \includegraphics[width=0.23\linewidth,clip,trim=7.2cm 5.5cm 1.7cm 3.5cm,cfbox=C3 3pt 0pt]{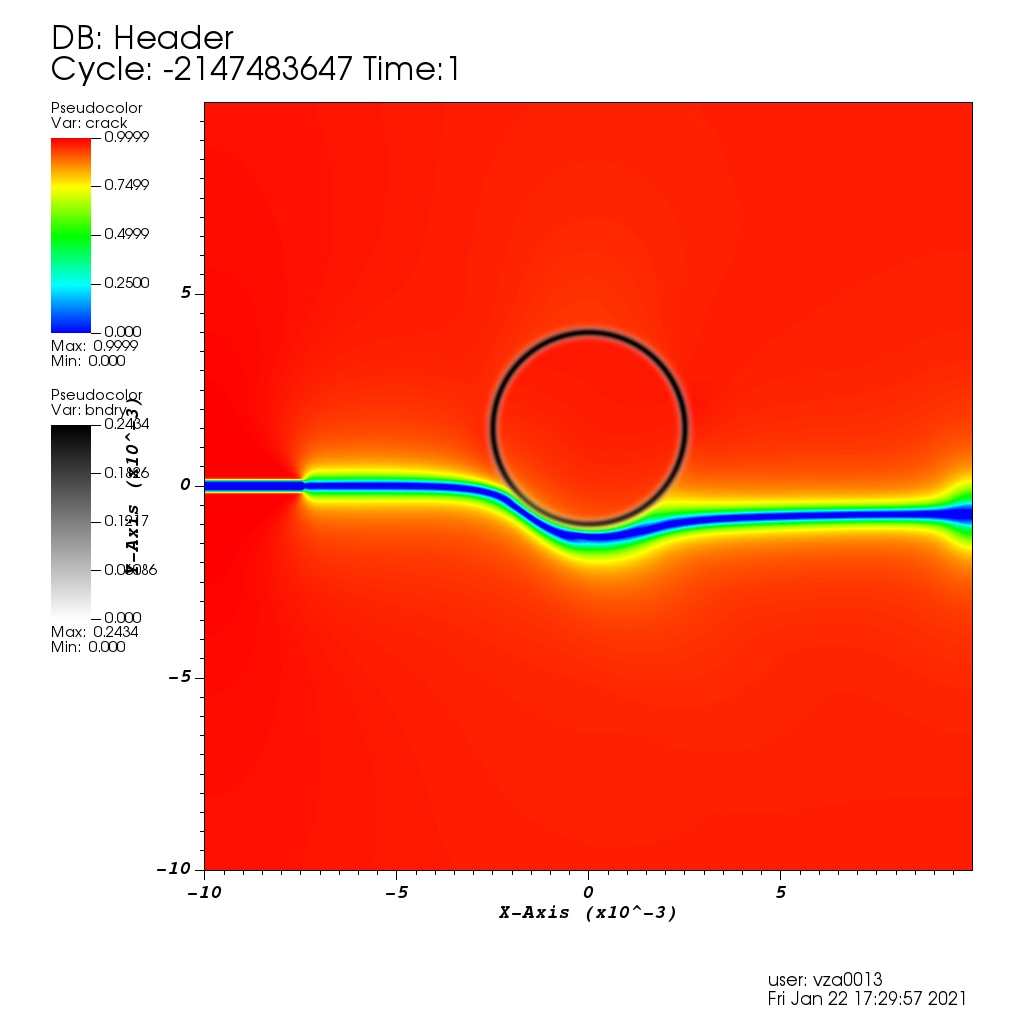}
    \includegraphics[width=0.23\linewidth,clip,trim=7.2cm 5.5cm 1.7cm 3.5cm,cfbox=C0 3pt 0pt]{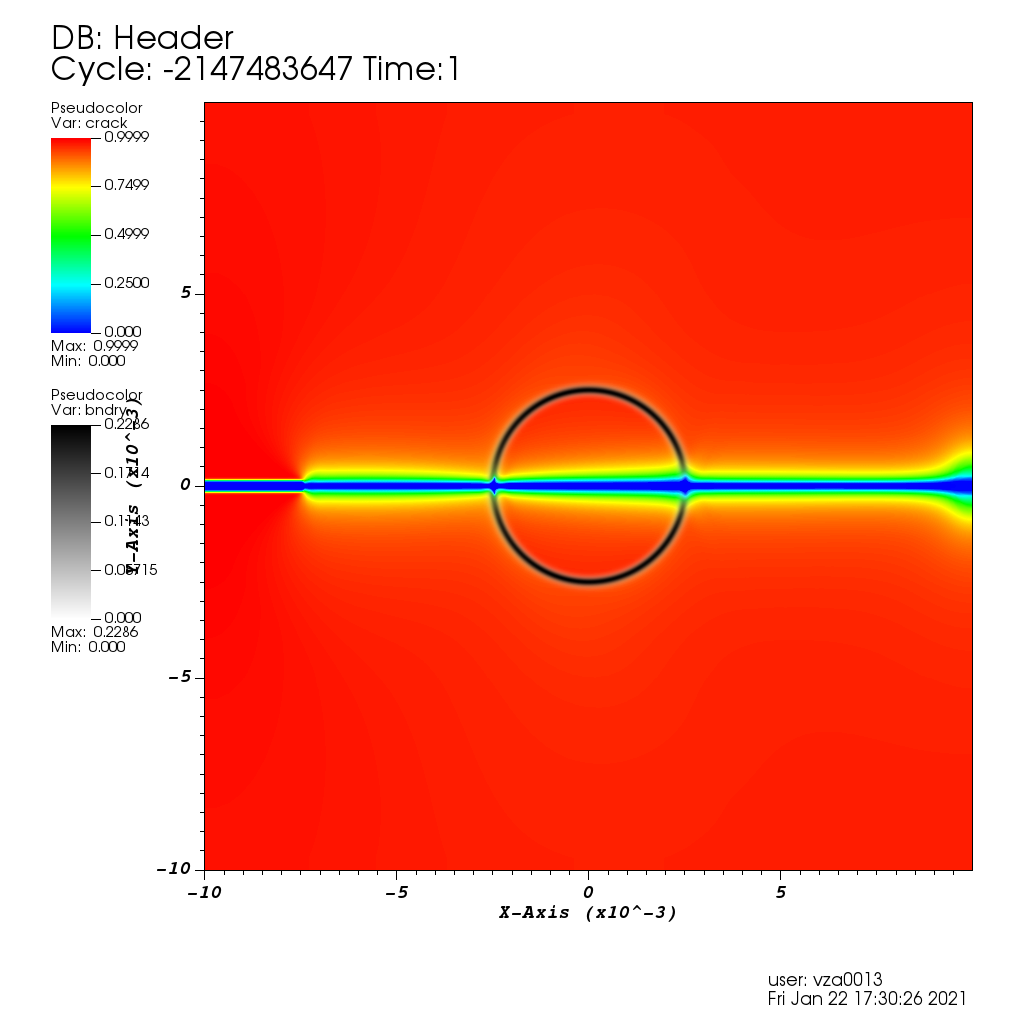}
    \includegraphics[width=0.23\linewidth,clip,trim=7.2cm 5.5cm 1.7cm 3.5cm,cfbox=C3 3pt 0pt]{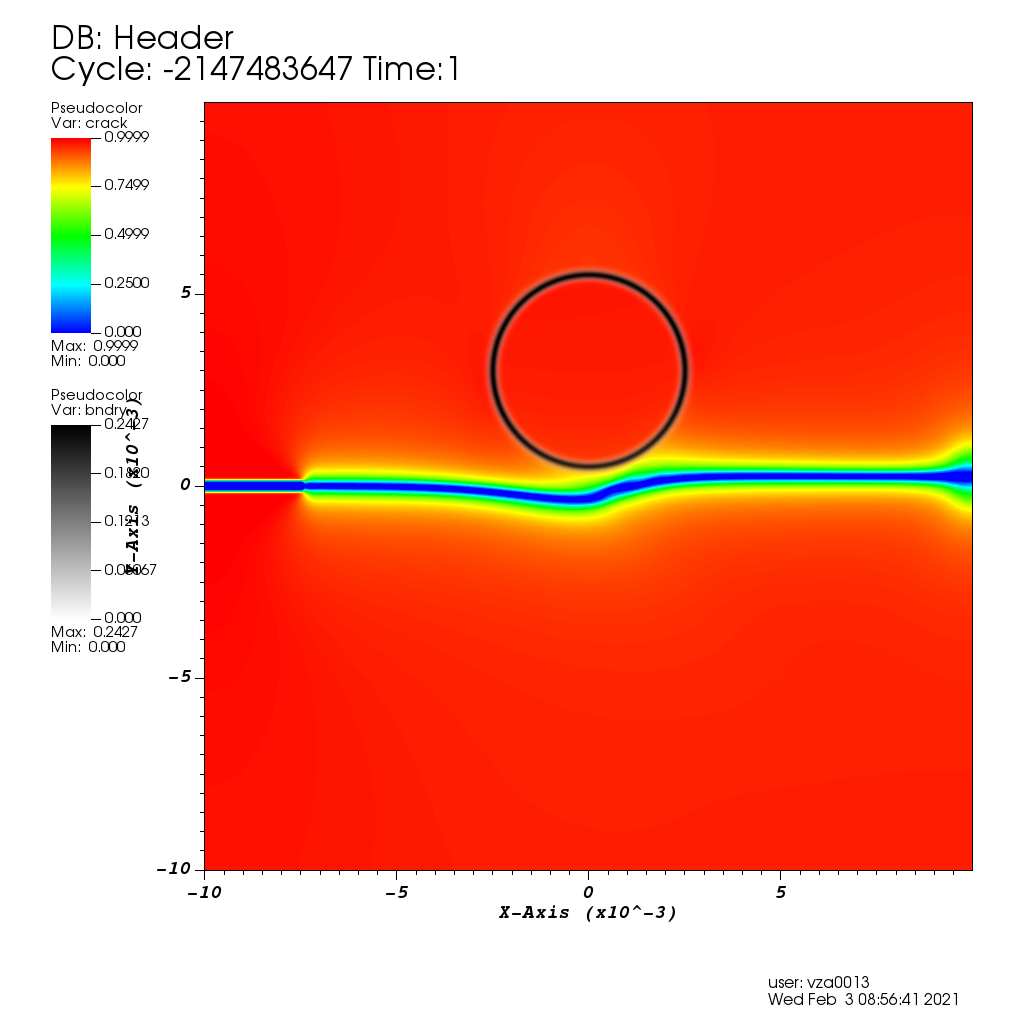}
  \end{minipage}
    
  \vspace{10pt}
  \begin{minipage}{0.05\linewidth}
    \ 
  \end{minipage}
  \begin{subfigure}[t]{0.95\linewidth}
    \begin{minipage}{0.23\linewidth}\caption{$\frac{G_{cd}}{G_{cf}}=0.75$, $\frac{h}{R}=0.6$}\end{minipage}\hfill
    \begin{minipage}{0.23\linewidth}\caption{$\frac{G_{cd}}{G_{cf}}=1.0$, $\frac{h}{R}=0.6$}\end{minipage}\hfill
    \begin{minipage}{0.23\linewidth}\caption{$\frac{G_{cd}}{G_{cf}}=0.65$, $\frac{h}{R}=0$}\end{minipage}\hfill
    \begin{minipage}{0.23\linewidth}\caption{$\frac{G_{cd}}{G_{cf}}=0.85$, $\frac{h}{R}=1.2$}\end{minipage}
  \end{subfigure}
  \caption{
    Inclusion: 
    Visualization of a selection of crack behavior.
    In the grid of images, each column of three images corresponds to a point on Figure~\ref{fig:crack_deflection}.
    The top image corresponds to a modulus ratio of $0.5$ (soft inclusion), the middle to a modulus ratio of $1.0$ (same material), and the bottom to a modulus ratio of $2.0$ (hard inclusion).
    The box outline color corresponds to the categorization ({\bf\color{C0}inclusion fracture}, {\bf\color{C1}branching}, {\bf\color{C2} delamination}, {\bf\color{C3} matrix fracture}).
  }
  \label{fig:VisualizationDeflection}
\end{figure}

The modulus ratio $E_i/E_m$ plays a critical role in governing crack behavior. 
Changing modulus ratio introduces complex stress distribution around the inclusion causing cracks to deflect. 
Increasing the modulus ratio from $0.5$ (soft inclusion) to $1.0$ (same material) to $2.0$ (hard inclusion) can result in crack changing from inclusion fracture to delamination to matrix fracture depending on $G_{cd}/G_{cf}$ and offset ratios.
Figure \ref{fig:VisualizationDeflection} shows actual crack paths for $12$ cases with different modulus ratios. 
Column (a) of Figure \ref{fig:VisualizationDeflection} shows one such case where the crack deflects ``up'' to cause inclusion fracture for  $E_i/E_m = 0.5$, and deflects ``down'' to cause matrix fracture for $E_i/E_m = 2.0$.
In such cases, the energetic cost associated with deflection is lower than the overall cost of propagation.
The effect of modulus ratio is also evident in column (d) of Figure \ref{fig:VisualizationDeflection} where the crack deflects ``up'' for $E_i/E_m = 0.5$ (inclusion fracture) and $2.0$ (matrix fracture), but not for $E_i/E_m = 1.0$ (matrix fracture).
This is because for $E_i/E_m = 1.0$, there is no stress concentration around the inclusion.
This, in conjunction with high offset and higher $G_{cd}/G_{cf}$, results in a classic Mode-I type matrix fracture.

\begin{figure}
  \centering 
  \includegraphics[width=\linewidth]{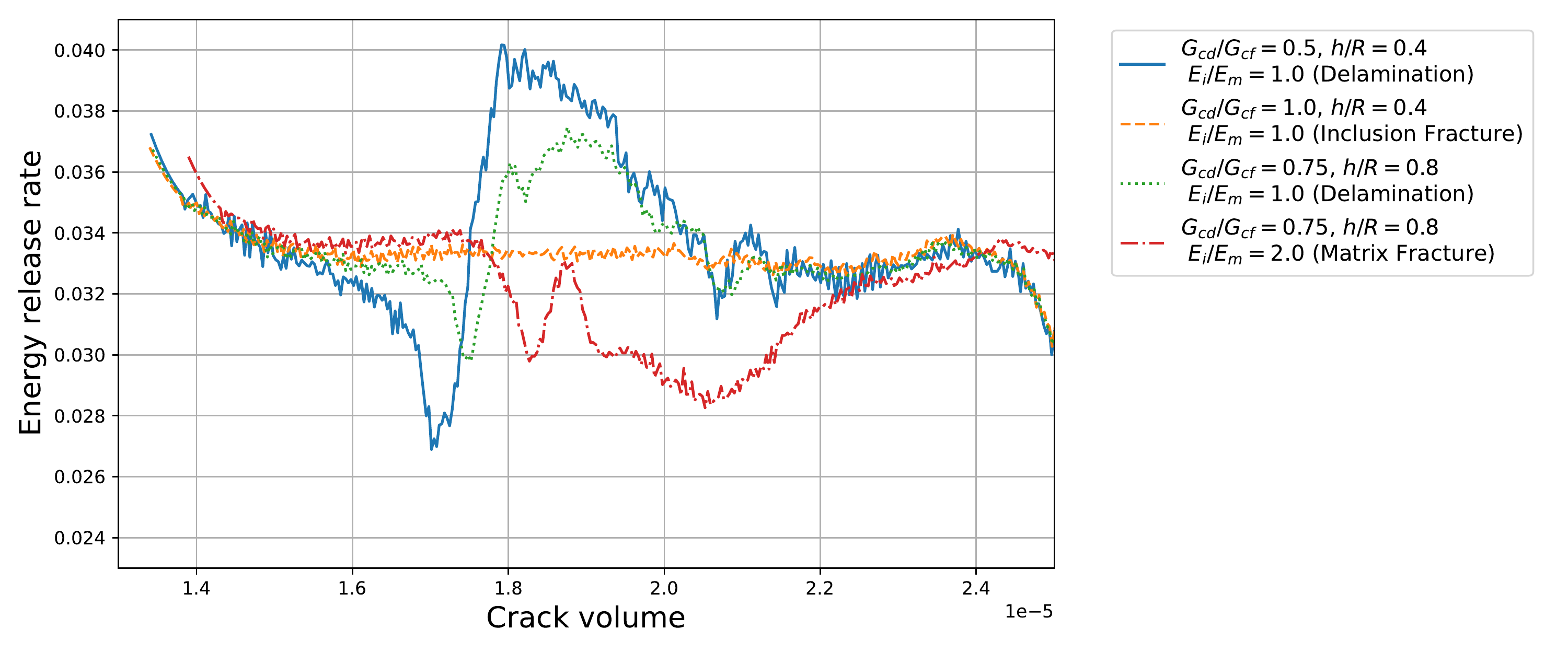}
  \caption{
    Inclusion: Evolution of crack energy release rate as it propagates through the domain.
  }
  \label{fig:Deflection_ERR}
\end{figure}

Figure \ref{fig:Deflection_ERR} shows the evolution of the fracture energy release rate $G$ as the crack propagates through the domain. 
A baseline response of flat $G$ vs crack curve is obtained (orange, large dashes) for $G_{cd}/G_{cf} = 1.0$, $h/R = 0.4$ and $E_i/E_m = 1.0$ corresponding to Mode-I type fracture.
For the rest of the cases, the $G$ curves are outcome of complex stress states and energy distribution around the crack and the inclusion.
For cases other than the baseline, the $G$ curve exhibits a sharp dip corresponding to the crack being in the vicinity of the interface.
For delamination cases (solid blue and dotted green), $G$ first shows a dip and then shows an increase above the baseline Mode-I type response.
For the matrix fracture case (dash-dot purple), the crack deflects down to avoid the inclusion resulting in a $G$ curve that mostly stays below the baseline response.

\section{Conclusions}

The contributions of this work are twofold: First, we have presented a novel framework for efficiently solving phase field fracture problems using block-structured adaptive mesh refinement.
Second, we have demonstrated the efficacy of this framework by investigating a large set of problems in heterogeneous fracture mechanics.

The proposed method has demonstrated efficiency and accuracy, but is not without a number of limitations.
Like most phase field fracture codes, solving the equations of elasticity requires a high number of solver iterations due to the poor condition number of the elastic operator for the degraded material.
It should also be pointed out that the performance of the method generally decreases as the crack grows in length, as the entire crack length must be fully resolved at all times during the simulation.
\added[id=R2,comment={2.2}]{(We note that by modifying the refinement criterion and the elastic solver, it may be possible to coarsen the non-evolving portion of the crack and avoid this issue; however, this presents other challenges that would need to be addressed in future work.) }
The use of BSAMR with a regular grid also means that all non-rectangular geometry must be included in a diffuse-interface manner.
Finally, the use of a rectangular grid limits the applicability of the method.
While it has a wide range of application to representative volume element (RVE) type simulations, it has limited applicability to non-rectangular domains.

Three types of heterogeneous structures were examined using the proposed method.
A large suite of simulations was completed for each type of structure.
(We note that not all simulations ran to completion; however, in each case, simulations had already progressed far enough to determine the crack propagation behavior. )
First, laminate-type structures were investigated to determine the regions of crack behavior with respect to the delamination vs fracture energy release rate, and the angle of the interface.
The results were compared with an analytic (albeit simplified) model, and reasonable agreement was determined.
Second, the case of a sinusoidal interface was considered, with analysis similar to the above but with respect to the amplitude of the interface rather than the angle. 
As with the first case, reasonable agreement was observed between the analytic estimation and the results of the parameter study.
Finally, the case of a circular inclusion was considered, where the stiffness of the conclusion varied from $1/2$ to $2$ times that of the matrix.
In addition to branching, fracture, and delamination, crack splitting and deflection was also observed.

\section*{Acknowledgements}

B Runnels was funded by the Office of Naval Research, grant number N00014-21-1-2113.
The authors are grateful for the support of the Auburn University Easley (2020) Cluster for assistance with this work.

\appendix

\section{Multi-level operator consistency}\protect\label{sec:multi_level_consistency}

Bijective numerical operators $D_n$ (operating on level $n$) in the proposed framework must be implemented such that the following is maintained:
\begin{align}
    D_n = R\circ D_{n+1} \circ I\label{eq:appendix_opconsistency}
\end{align}
where $R:C(\Omega_{n+1})\to C(\Omega_n)$ is the surjective restriction operator and $I:C(\Omega_{n})\to C(\Omega_n)$ the injective interpolation operator.
Failure to implement $D_{n}$ such that the above consistency relation is maintained results in error accumulation and convergence problems.

In the case of $D_{n}=id$, the identity operator, one may define restriction/interpolation operators that $(R\circ I)(f)=f$ $\forall f \in C(\Omega_n)$; that is, $R$ is defined to be the inverse of $I$ over $\operatorname{image}(I)\subset C(\Omega_{n+1})$.
Then $id$ trivially satisfies (\ref{eq:appendix_opconsistency}).
However the converse is not true in general.
That is, if $D_{n}$ is bijective, $I$ is strictly injective, and $R$ is strictly surjective, then
\begin{equation}D_{n+1} \ne I \circ D_{n} \circ R.\end{equation}
This may be seen by considering an $f \in C(\Omega_{n+1}) \setminus \operatorname{image}(I)$.
Then $R(f)\in C(\Omega_{n})$, as is $D_n(R(f))$.
But, by definition, $I(D_n(R(f)))\in\operatorname{image}(I)$.
Consequently, since $\operatorname{image}(I\circ D_n\circ R)\subset \operatorname{domain}(D_{n+1})$, the restricted operator is non-bijective and therefore not equal to $D_{n+1}$.

\begin{figure}
    \includegraphics[width=\linewidth]{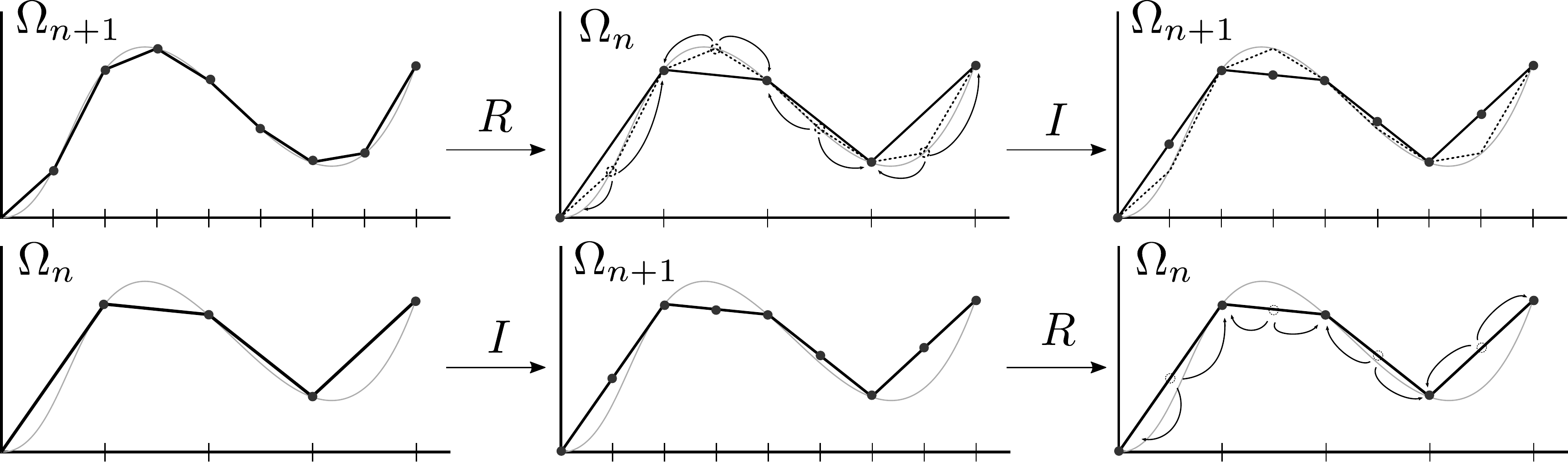}
    \caption{(Top) A function on $\Omega_{n+1}$ subject to $I\circ R$ (Bottom) A function on $\Omega_n$ subject to $R\circ I$.}
    \label{fig:bijective}
\end{figure}

This may also be understood graphically (Figure~\ref{fig:bijective}).
A function on the fine level ($\Omega_{n+1}$) experiences an irreversible loss of information when restricted to the coarse level, which cannot be restored upon interpolation back to the fine level.
On the other hand, since no new information is introduced when a function on the coarse level is interpolated to the fine level, there is no loss of new information when it is restricted back to the coarse level.

\bibliographystyle{ieeetr}
\bibliography{library}

\end{document}